\documentclass[11pt,preprint]{aastex}

\usepackage{lscape}

\def\stacksymbols #1#2#3#4{\def\theguybelow{#2}
        \def\verticalposition{\lower#3pt}
        \def\spacingwithinsymbol{\baselineskip0pt\lineskip#4pt}
        \mathrel{\mathpalette\intermediary#1}}
\def\intermediary #1#2{\verticalposition\vbox{\spacingwithinsymbol
        \everycr={}\tabskip0pt
        \halign{$\mathsurround0pt#1\hfil##\hfil$\crcr#2\crcr
                \theguybelow\crcr}}}

\def\gta{\stacksymbols{>}{\sim}{3}{.5}}

\begin{document}

\title{Cold Dust in Early-Type Galaxies. I.  Observations\altaffilmark{1}}

\author{Pasquale Temi}
\affil{Astrophysics Branch, NASA/Ames Research Center, MS 245-6,
Moffett Field, CA 94035; \\ \&  SETI Institute, Mountain View, CA 94043.}

\author{Fabrizio Brighenti}
\affil{Dipartimento di Astronomia,
Universit\`a di Bologna, via Ranzani 1, Bologna 40127, Italy;\\  \&
University of California Observatories/Lick Observatory,
Board of Studies in Astronomy and Astrophysics,
University of California, Santa Cruz, CA 95064.}

\author{William G. Mathews}
\affil{University of California Observatories/Lick Observatory,
Board of Studies in Astronomy and Astrophysics,
University of California, Santa Cruz, CA 95064.}

\author{Jesse D. Bregman }
\affil{Astrophysics Branch, NASA/Ames Research Center, MS 245-6,
Moffett Field, CA 94035.}

\altaffiltext{1}{Based on observations with ISO, an ESA project
with instruments funded by ESA Member States (especially the PI countries:
France, Germany, the Netherlands and United Kingdom) and with the
partecipation ISAS and NASA}

\begin{abstract}
We describe far infrared observations of early-type 
galaxies selected from the ISO archive. 
This rather inhomogeneous sample includes 39 giant elliptical 
galaxies and 14 S0 (or later) galaxies. 
These galaxies were observed with the array photometer 
PHOT on-board the ISO satellite using a variety of different 
observing modes -- sparse maps, mini-maps, over-sampled maps 
and single pointings -- 
each of which requires different and often rather 
elaborate photometric reduction procedures. 
The ISO background data agrees well with COBE-DIRBE results
to which we have renormalized our calibrations. 
As a further check,
the ISO fluxes from galaxies at 60 and 100 $\mu$m 
agree very well with those previously observed with IRAS 
at these wavelengths.
The spatial resolution of ISO is several times greater than 
that of IRAS and the ISO observations 
extend out to 200 $\mu$m which views a significantly greater mass 
of colder dust not assessable to IRAS.
Most of the galaxies are essentially point sources at ISO 
resolution, but a few are clearly extended at FIR wavelengths 
with image sizes that increase with FIR wavelength. 
The integrated far infrared luminosities do not correlate with 
optical luminosities, suggesting that the dust may have 
an external, merger-related origin. 
In general the far infrared spectral energy distributions can be 
modeled with dust at two temperatures, 
$\sim 43$ K and $\sim 20$ K, which probably represent 
limits of a continuous range of temperatures.
The colder dust component dominates the total mass of dust, 
$10^6 - 10^7$ $M_{\odot}$, 
which is typically more than ten times larger than 
the dust masses previously estimated for the same galaxies 
using IRAS observations. 
For S0 galaxies we find that 
the optically normalized far-infrared luminosity 
$L_{FIR} / L_B$ correlates strongly with the mid-infrared luminosity
$L_{15\mu\rm{m}} / L_B$, but that correlation 
is weaker for elliptical galaxies.

\end{abstract}


\section{Introduction}

Most of our current knowledge of the infrared emission from elliptical
galaxies is based on data from the {\it Infrared Astronomical 
Satellite} (IRAS) at 12, 25, 60 and
100$\mu$m \citep{kna89}.
The IRAS satellite detected emission at all of these
wavelengths from many bright elliptical galaxies, verifying the presence of
dust in these galaxies having a wide variety of temperatures. 
Unfortunately, the spatial resolution of IRAS
observations, 3 - 5\arcmin~ at 100 $\mu$m, is much larger than the
optical size of most elliptical galaxies. 
Consequently, it has been
impossible to disentangle various possible sources of dust emission. 
IRAS fluxes may confuse and combine emission from AGN, central
dust clouds, circumstellar dust around red giant stars and 
truly interstellar dust
distributed throughout the galactic volume.
Worse, the large IRAS aperture may include irrelevant 
emission from other nearby sources.

Fortunately, 
the {\it Infrared Space Observatory} (ISO) \citep{kes96} has also detected 
many early-type galaxies at FIR wavelengths where emission 
from cold dust is expected to have its peak.
The higher spatial resolution afforded by ISO is a necessary and
essential attribute for a proper interpretation of infrared data from
elliptical galaxies.  
ISOPHOT, the diffraction-limited 
imaging photometer on-board the ISO satellite \citep{lem96}, also has 
a fainter limiting magnitude compared to IRAS and extends to somewhat 
longer wavelengths, 200$\mu$m, permitting
a better match to the cold 
dust emission expected from elliptical galaxies. 

The origin of dust in elliptical galaxies 
is controversial. 
One confirmed source of dust 
is the collective outflow of dusty gas from evolving 
red giant stars. 
Emission attributed to circumstellar dust in E galaxies 
has been detected at $\sim 10$ $\mu$m by 
\citet{kna89} and \citet{ath02}.
A circumstellar origin seems secure 
since this emission shares the same 
de Vaucouleurs profiles with the optical images. 
As dusty gas moves away from orbiting red giants,
it is eventually decelerated against the 
local hot interstellar gas,
$T \sim 10^7$ K, and subsequently undergoes a variety of 
complex gasdynamical interactions and instabilities.
If the stellar dust grains ultimately come into contact with 
the hot interstellar gas, they would be 
eroded (sputtered) by thermal collisions with ions and destroyed 
in only $\sim 10^7 - 10^8$ years.
Nevertheless, during their relatively short lifetimes the grains 
absorb energy from starlight and incident thermal 
electrons and convert this energy to radiation 
mostly at far infrared wavelengths 
(e.g. Tsai \& Mathews 1995, 1996; \citet{tem03}). 
In this type of dust evolution we would expect that the metals
contained in the dust grains would eventually merge with 
the hot interstellar gas and produce some of 
the observed X-ray line emission.

But this simple model for dust evolution does not 
take into consideration the highly efficient cooling 
that occurs when thermal electrons lose energy 
by inelastic collisions with dust grains.
\citet{mat03} have recently argued that 
dusty gas ejected from stars 
may cool before or soon after entering the hot gas phase. 
The cooling rate of the hot dusty gas 
is dominated by emission from dust grains, 
not thermal X-ray emission. 
If this is the dominant evolutionary path for stellar dust, 
the thermal energy of the cooling ambient gas 
is emitted by the dust as far infrared emission.  
This type of rapid dust-induced cooling may create the 
dusty disks and clouds seen in the majority 
of luminous elliptical galaxies (e.g., \cite{van95,fer99}). 

While stellar winds are a certain source of dust in elliptical 
galaxies, many authors have suggested that mergers with 
dust-rich dwarf galaxies are in fact 
the dominant source of dust.
One good argument in favor of an external source of dust 
in elliptical galaxies is the 
lack of any strong correlation between FIR and optical 
galactic luminosities
(e.g. \citet{for91}; \citet{gou95}; \citet{tri02}).
Further support for an external origin comes from 
counter-rotating warm gas observed in some elliptical 
galaxies (e.g. \citet{cao00}).

This is the first of a series of papers in which we 
shall explore and 
interpret archival ISO observations of early-type galaxies.
In this paper we review the ISO observational data
and describe the data reduction analysis used to retrieve the spatial
location and emission spectrum of dust in these galaxies.
In a subsequent paper we will use these data
to investigate the origin, evolution, and physical properties of dust
in early-type galaxies.

\section {The sample}

Working under the ISO Guest Investigator Programme, 
we searched the ISO archive to extract FIR data
($\lambda \ge 60$ $\mu$m) for all the early-type galaxies
observed with {\it ISOPHOT}. We excluded M86 from the sample
because it has already been studied in detail by Stickel et al. (2003).
No other selection criteria were used.
All the far-IR data presented in this paper are unpublished 
except the data relative to NGC 3998 \citep{kna96}.
As expected from archival data, 
the sample is varied and somewhat inhomogeneous.
Table 1 shows a detailed log of the ISOPHOT observations 
that we reduced and analyzed; entries in the
table include for each source the coordinates, 
the ISO observation time 
identification number (TDT), the filter used, the observing mode,
the field of view, the raster size when applicable and the 
total integration time.
While most of the target galaxies in our sample have full coverage
in wavelength from 60 to 200 $\mu$m,
some do not, depending on the particular scientific intentions of the
original ISO PIs who requested the data.

The relevant properties of the sample galaxies 
are summarized in Table 2.
The optical data are taken from the Lyon-Meudon
Extragalactic Database (LEDA; \citet{pat97})
and the Revised Shapley-Ames (RSA) Catalog \citep{san81}.
Of the 58 galaxies in Table 2, 43 are ellipticals and 
15 are S0 or later types, according 
to RSA classifications.
Most objects in the sample are optically luminous 
``giant'' galaxies, which are our primary targets 
for this investigation. 
Table 2 also includes a 
few early-type dwarf galaxies (NGC 147, NGC 185, NGC 221,
NGC 2328), defined here as galaxies with $L_B \le 10^9$ L$_{B,\odot}$,
that have been observed by ISO, 
but we do not include these galaxies in our final discussion.
Since the evolution of interstellar dust 
in giant elliptical galaxies may be strongly influenced
by its interaction with a hot interstellar medium (Tsai \& Mathews
1995, 1996; \citet{tem03}), we also list in Table 2 the
X-Ray luminosity $L_x$, taken from \citet{osu01}, which is 
a measure of the amount of hot gas present in each galaxy. 
Only an upper limit to $L_x$ is available for 24 objects.
As expected from the well-known 
large scatter in the $L_B - L_x$ relation for elliptical
galaxies (e.g. \citet{esk95}), the galaxies in our sample
also span a wide range of $L_x$. 
A few galaxies in our sample are fairly strong radio sources
and many may harbor an AGN or have LINER optical spectra. 
As a measure of the non-thermal emission from these galaxies, 
we list in column 5 of Table 2 
the 6 cm radio emission 
taken from the compilation of \citet{rob91}.
\citet{goud94} argues that 
the FIR emission from the vast majority of early-type galaxies 
is produced by cool dust, although a significant additional 
non-thermal component may be present in few objects
(e.g., NGC 4261 and NGC 4494).
To investigate the possible connection between the FIR emission and
the presence of cold gas, we also quote the HI mass (or upper limit) 
taken mainly from \citet{rob91} and the molecular ISM mass, 
both corrected for our assumed distance.
In the final column of 
Table 2 we list some very schematic notes
for each galaxy taken from the literature regarding 
the detection of dust in optical studies, the presence of 
warm emission line 
gas and the possible interaction of the 
galaxy with other nearby galaxies.

\section{Observations }

Of the large number of observing modes 
available for the ISOPHOT instrument \citep{kla02}, 
most of the nearly 60 early-type galaxies in the ISO archive 
have be observed with 
the photometric far-infrared cameras, C100 and C200, 
using observing configurations described by 
PHT 37/39, PHT 22, and PHT 32, 
referred to as the Astronomical Observing Template (AOT).
The C100 camera is used to observe at  60, 90, and 100 $\mu$m has a 
$3\times3$ pixel format with a $46^{\prime\prime}$ pixel scale.
The C200 camera used for observations at 150, 180 and 200 $\mu$m and 
has a $2\times2$ pixel format with a $92^{\prime\prime}$ pixel scale.

Our ISO sample is 
the largest currently available sample 
of nearby early-type galaxies studied 
in the FIR up to 200 $\mu$m.
These observations,
collected from scientific projects originally proposed by several PIs,
certainly do not form a complete or homogeneous sample.
Nevertheless, they have some statistical relevance and their maps represent
the deepest FIR images acquired to date, achieving sensitivities an order
of magnitude deeper than IRAS at 60 and 100 $\mu$m.
Also, ISOPHOT observations
taken in the oversampled map mode provide a spatial resolution far
greater then the previous IRAS data.
In a few cases the galaxies are resolved on scales comparable to
their optical images,
allowing a more trustworthy interpretation of their FIR emission.

The PHT 37/39 observational mode (AOT) produces sparse maps. 
These data were recorded in the
on-off mode, where observations were made in each filter with one 
on-source single point staring mode and one off-source exposure. 
Eight galaxies in our sample have been observed in this observing 
mode. The 60 and 100$\mu$m filters were observed with C100
and 180$\mu$m with C200.
Unfortunately, 
because of the limited field of view in the single pointing mode, 
some of the needed spatial information may be 
lost using this observing mode,  
but we were nevertheless able to perform 
some useful integrated photometry on the array with this AOT  
once the off-source has been subtracted. 

A large fraction of the sample has been observed 
in the PHT 22 AOT, either 
in the mini-map configuration or in the on-off chopping mode.
The PHT 22 mini-maps were made with a sequence of staring 
observations 
on a bidimensional regular spatial 
grid determined by spacecraft pointing.
Data were recorded at each pointing position with the 
same detector settings.
Eight galaxies were observed in the mini-map mode.
All but one have been observed at 60, 90, 170, and
200 $\mu$m with raster sizes of $3 \times 3$ pixels for C100
and $4 \times 2$ pixels for C200 in the $Y \times Z$ direction.
The final maps have sizes of $4^{\prime} \times 4^{\prime}$
and $7.5^{\prime} \times 4.5^{\prime}$, respectively.
For NGC1275 the only data available have
larger raster maps ($4 \times 4$ ) at 170 and 200 $\mu$m.
All measurements
have been recorded with an integral number of pixel offsets, 
i.e., multiples of offsets by 
$46^{\prime\prime}$ for the C100 array and 
$92^{\prime\prime}$ for the C200 array. 
Of the recorded eight galaxies, only three were detected but these 
appear as point-like sources with no distributed galactic emission  
at the sensitivity level of the instrument. 

Observations of 27 elliptical galaxies were made in the
single pointing, multi-filter mode PHT 22.
During each pointing, the focal plane chopper was used
to alternate between exposures of the source and a reference
field. Operating in a rectangular mode, the chopper has a throw
of $180^{\prime \prime}$ in the spacecraft Y-direction for both
C100 and C200 cameras.
Data was taken in four broad band filters
centered at 60, 90, 170, and 200 $\mu$m.
The integration times for measurements at 60 $\mu$m
were 64 sec on-source and
64 sec off-source in four chopper cycles, while in the remaining 
three bands the time spent on and off source was 32 sec each.
After the off-source subtraction,
the photometry is acquired by summing the flux in the pixels.

The PHT 32 AOT observing mode uses a
combination of raster mapping and chopper sweeping to create an
oversampled map.
These observations consist of a scan of $N \times M$ steps in the
Y and Z spacecraft coordinates respectively.
At each spacecraft pointing position the focal plane
chopper samples the sky in steps of $1/3$ of the pixel size, which is
$15^{\prime\prime}$ at 60 and 100 $\mu$m and $30^{\prime\prime}$
at 150 and 200 $\mu$m. The raster sampling interval in Y is set to a multiple
of the chopper sampling interval, while the sampling in the Z direction is
controlled entirely by the spacecraft raster sampling interval in Z.
Depending on the step intervals chosen to construct the raster map,
the linear scans from each detector pixel can be combined
to achieve a sampling in the sky that approaches the Nyquist limit
$\Delta\Theta = 17^{\prime\prime}$
at 100 $\mu$m.
Using either the C100 or C200 array cameras,
PHT 32 observations can provide a sampling of
$15^{\prime\prime}\times 23^{\prime\prime}$
and $30^{\prime\prime}\times 46^{\prime\prime}$ over the sampled field,
respectively.
Typically, the maps presented here were made with raster
sizes large enough to detect, if present,
any faint emission due to cold dust distributed throughout the galaxy.
In addition, as the dimension of the spacecraft raster
in the Z direction is increased,
the on-source integration time also increases
since one row of the C100 or C200 detector continuously scans
through the source, thus sampling the same region multiple times.
These large maps
also help determine the appropriate flat field correction. 

Recently many ISOCAM observations 
of mid-infrared dust emission from early-type galaxies
have been published  
(\citet{fer02}, \citet{ath02}, \citet{mal00}).
While we concentrate here on the far-infrared emission from 
cold dust, it is of interest to find possible correlations
between L$_{MIR}$ and  L$_{FIR}$ for galaxies detected in both bands. 
Many of the galaxies in our sample have been observed 
in staring mode with ISOCAM at mid-infrared wavelengths 
(CAM01 AOT). 
The images have been taken with the 32$\times$32 Si:Ga array camera 
at the pixel scale of $6^{\prime \prime}$ in the LW3
($\lambda_{ref}=14.3 \mu m$), LW9 ($\lambda_{ref}=14.9 \mu m$),
and LW10 ($\lambda_{ref}=12.0 \mu m$) filters. The LW10
filter has been designed to mimic the IRAS 12 $\mu$m filter
bandwidth. LW3 and LW9 have similar central wavelength, but a
$\lambda$  / $\delta \lambda$ = 3 and 8, respectively.
Each observation consists of a collection of exposures,
each one taken with the target galaxy centered at several positions 
on the array.
Two galaxies, NGC 4261 and NGC4636, have been 
observed in the MIR 
with a plate scale of $3^{\prime \prime} / pix$, while 
large raster maps have been made for NGC 4552 and IC 1459. 
In this paper we reduced and obtained photometry at 15 $\mu$m for a 
dozen early-type galaxies that have been detected by ISOPHOT.

\section{ Data Reduction}
Although all data presented here have been recorded using only the two 
C100 and C200 far-infrared cameras, 
the data reduction process has to be specifically 
tailored for each AOT observing mode. 
In particular the detector response and flux 
calibration  can be significantly 
different depending on the specific AOT employed.
In the following we describe in some detail the 
data reduction procedure for each specific observing mode.

\subsection{PHT 37/39 AOT }

Data reduction and calibration
were carried out using 
the ISOPHOT Interactive Analysis package (PIA) \citep{gab97},
together with the calibration data set V4.0
\citep{lau98}. The reduction
included corrections for non-linearity, cosmic
particle hits, and linear 
fitting of the signal ramps. After resetting all ramp
slopes and subtracting the dark current, the detector
responsivity was calibrated using measurements of the
thermal fine calibration source (FCS) on-board.
The sky-subtracted source signal 
was then corrected for the fraction of the point-spread
function not included in the field of view of the detectors. 
The error propagation in ISOPHOT
data reduction is described by \citet{lau99}. The statistical
errors derived from signal processing are about 5-20 percent, 
depending on
the wavelength range and object brightness, but systematic errors due
to absolute calibration accuracy 
are estimated to be 30 percent \citep{kla98}.
As reported in 
the notes of Table 2, two galaxies, NGC 4105 and NGC 5353, appear 
to have companions that may be partially included in the beam 
and these companions may contribute 
to the detected flux (see below). 
Two galaxies, NGC 4291 and NGC 6876, were not detected at any of the
three bands (60, 100, and 180 $\mu$m) observed.

\subsection{PHT 22 AOT mini-maps }

All data have been processed using OLP version 10.1 
and the reduction
has been performed with PIA version 10.0. 
Data were corrected for non-linearity and cosmic
particle hits. A linear fit is then applied 
to the ramps providing the signal per ramp data. 
A second  cosmic ray hits correction was performed at this stage.
Then all ramp signals are averaged over time at each raster
point. At this stage the data are corrected for reset interval
and the dark current is subtracted. A signal linearization
is then applied to the data to correct for a changing response
during the measurements.

The same reduction procedure is applied to the complementary 
measurements of the thermal FCS used 
to calibrate the fluxes. Two FCS observations are associated with each
mini-map data set; one is recorded
before and one after the source observation. An average of the two FCS 
measurements is converted to a real flux using tables constructed from sources
for which the flux is well determined. 
After calibration the data are ready for mapping and the flux densities 
can be extracted.

Most of the galaxies in our sample are faint and, when detected, 
appear as point sources in the reduced maps. 
Uncertainties in the flux calibrations introduce errors of the order of 
30 percent in the derived flux densities. 
The detection of a source, however, does not depend on the performance
of the flux calibration, and the S/N ratio
should be estimated before the
flux calibration is performed. 
The raster maps constructed for these observations are well suited
to derive the photometry of a point source. The central position is 
sampled nine times but enough data is available for 
background positions to make an accurate background subtraction.

Figure 1 shows the layout of the raster maps on the sky.
Our observations have been constructed with raster steps equal
to an integral number of pixels. This facilitates the derivation of 
flux densities from the maps since the 
correction factors for the point-spread function 
are only available for 
one-pixel intervals in both PHOT-C cameras. 
Under these conditions we are able to determine fluxes for
each individual pixel. 
For both arrays the PSF is larger then one pixel, implying that each 
pixel at each raster position collects some emission 
from the central source plus background. 
The fraction of the central source flux
allocated to each pixel in the array 
has been estimated empirically by \citet{Lau99}. 
Once the PSF correction is applied, 
the source and background flux densities can be extracted. 
Details concerning this reduction algorithm can be found in \citet{ric02}
and \citet{kla00}. 

\subsection{PHT 22 AOT single-pointing}

For single-pointing observations in the PHT 22 AOT
we used PIA version 10.0,
which includes all steps needed to fully process the raw data.
Since this procedure is similar to our reduction of  
the PHT 22 mini-maps, we emphasize here 
the differences in the reduction process required in the single-pointing mode. 

After correcting for non-linearity and cosmic particle hits,
the on and off signals can be derived from the chopped measurements 
using the ``Ramp Pattern" processing algorithm that uses a calibration
table to correct for signal 
losses due to chopping. 
The first and last
few readouts of each ramp were discarded since 
they are often affected by spurious electronic effects.
Before deriving the signals, we inspected, and in some cases removed, 
part of the ramps that presented anomalies.
We always discarded the first signal 
for chopper plateau since the detector response was not stabilized. 
The calibration of detector responsivity was performed using the associated
measurement of the FCS. 
The source fluxes were corrected for transient
effects and signal losses due to detector hysteresis in the chop cycle 
and the effect of vignetting on the illumination of the detector array.
The accuracy of the absolute photometric calibration depends
on systematic errors \citep{lau98} and it is known to be better than
30 percent.
Typical 3$\sigma$ noise limits for chopped observations were around 
100 mJy and 150 mJy using the C100 and C200 detectors, respectively.
The noise limits for regions observed with high interstellar 
cirrus confusion were a factor 3 or 4 higher than those with low cirrus 
emission. 

\subsection{PHT 32 AOT}

These observations suffer from strong signal transients due to flux changes
generated by the relatively fast chopper movement. To properly correct
the data, we used procedures specifically developed for the PHT 32 AOTs
by the MPI Kernphysik (Heidelberg) and supported by
the ISO Data Center. These new procedures and routines have been recently
incorporated into the standard PIA package  \citep{gab97},
and are described in detail by \citet{tuf03}. 
Here we briefly outline the most important steps taken in reducing the PHT 32
observations.

The ISOPHOT C200 Ge:Ga stressed array detector operating between
120 and 200$\mu$m, and especially the C100 Ge:Ga array detector
operating between 60 and 100$\mu$m have complex non-linear responses.
The non-linearities depend on the
illumination history over timescales of 0.1-100 sec,
on the absolute flux level and on the flux changes involved.
Normally processed
ISOPHOT data can be reduced in one pass from the raw input data with full
time resolution to the final calibrated map. To correct for responsivity
drift effects, however, it is necessary to iterate between a sky map
and the input data at full time resolution, and this has been
implemented in the dedicated data reduction package.

The first basic step in reducing
the PHT 32 mapping observations is signal conditioning.
This procedure aims to provide a data timeline free of
artifacts such as glitches, dark current, and non-linearity effects,
but 
retains all knowledge of the transient response of the detector to the
illumination history at the full time resolution
of the input data.
Signal conditioning is applied individually to each detector pixel.
Also at this stage a rectangular
grid of pointing directions on the sky,
the {\it natural grid}, is defined with
information collected from the spacecraft pointing history during
the specific observation.
The natural grid,
defined by the combination of chopper sweeping and raster mapping,
has pixels that are $15^{\prime\prime}$ and $30^{\prime\prime}$ wide in the
spacecraft Y coordinate for the C100 and C200 array detectors, respectively.
This oversampling performed by the PHT 32 observing mode becomes very useful
when constructing maps of the surface brightness distribution of a sky
field since it
eliminates interpolation or any other map reconstructing
techniques that often introduce artifacts in the final map.
During signal conditioning, the data are corrected for nonlinearity effects
in the integration ramp and dark current subtraction.
A deglitching procedure is
also applied to remove spikes in detector responsivity from cosmic rays
hitting the array.

The second basic step in reducing PHT 32 maps is 
transient correction, an iterative process to determine the
most likely sky brightness distribution giving rise to the observed signals.
This is a nonlinear optimization problem. The sky brightness at each
point in the predefined natural grid is the variable that
is optimized iteratively by minimizing
the difference between the input data and a
combination of the detector response model with a trial data timeline.
Since no theory exists to predict the detector response behavior,
an empirical model has been found to apply the transient correction  to
the data. The severity of the data reduction difficulty
depends on both the source/background
ratio and the dwell time at each pointing direction. Because galaxies in our
sample are not very bright, the transient corrections are not as
large as those of bright compact sources,
but they are nevertheless crucial to achieve
the highest accuracy in the integrated photometry.

Before running the routines designed to reduce the PHT 32 
data (P32tools),
we processed the two FCS measurements taken before and after
each raster observation, using the standard reduction 
with PIA. At first, once the P32tools package is started,
the data are processed and corrected for ramps linearization,
dark current subtraction end the reset interval correction.
After this preliminary reduction the  process
become interactive and the chosen parameters are adjusted
to each single observation. 
For the drift fitting procedure, we used the internal dispersion
as the errors to be considered, and we included the slew data
in the ``coadded'' or ``sequence'' option depending on the 
brightness of the object. The number of iterations
of the starting condition and for the illumination history were
left to their default values of 1 and 2 for the C100 and C200
arrays, respectively. At first the drift correction was applied 
for all detector pixels with the default fitting parameters.
We inspected single pixels to see the need for modification of
the drift fit and also to apply a manual deglitching. 
Since most of the galaxies in the PHT 32 sample are faint, the
degliching procedure became critical and was applied 
regularly on a single pixel inspection basis.
When the inspection of drift corrected map was satisfactory, 
we exported the results of the P32tools package to PIA.
Once the transient correction has been applied to the observed signal,
we proceeded to construct the final calibrated maps.
For each galaxy we used the FCS measurements
to convert the signal
from detector units ($V \ \  s^{-1}$) to $mJy \ \ ster^{-1}$.
For each filter exposure,
we averaged the detector responses of the two FCS measurements. 
Finally we applied the flat field
correction assuming a uniform background emission.

\subsection{CAM 01 AOT }

Data reduction has been performed using the Camera Interactive
Analysis (CIA) package, version 5.0 \citep{ott97}. 
The procedure to reduce ISOCAM data starts with the dark current 
subtraction, for which we used the {\it model} option to take 
into account the long term drift in the dark current which 
occurred during the ISO mission. Then we removed the cosmic
ray hits using a sigma clipping filter or muli-resolution
method. This removes most of the short-term duration 
glitches. We also visually inspected the frames to manually
remove glitches with very long time constants that may affect
more then one frame. After the memory effects were corrected 
using the Fouks-Schubert method, we flat-fielded the data 
with library flat-fields. The conversion from electronic units
to Jy/pix was performed using the standard CAL-G calibration data
that accompanies the data products.
The sky background was evaluated by taking the average pixel value
from several blank regions in the reduced frames. 
After each single exposure 
has been properly reduced, the final image was produced by registering 
and coadding each frame.

\section {Calibration Accuracy}

In this section we compare the absolute calibrations
of the ISO fluxes with those observed with other FIR
instruments. 
The standard method of ISOPHOT flux calibration 
is to measure the fine calibration source (FCS), 
a greybody inside the ISOPHOT instrument. 
Using the mean of 
two FCS measurements recorded before and after each source 
observation, the  signals are converted to surface brightness
values in MJy $sr^{-1}$. 
The observed FCS flux is converted to a real flux 
using calibration tables constructed from observations of standard stars,
planets and asteroids.
We have cross-checked the ISO FCS-based calibrations  by 
comparing the surface brightnesses of the backgrounds measured
by C100 and C200 cameras with a background model based on 
COBE-DIRBE observations. 

The background is estimated at each galaxy position by combining the 
contributions from the zodiacal light and Galactic emission. 
The zodiacal light emission depends on the ecliptic latitude 
$\beta$ and the solar elongation $e$. Observing constraints for the 
ISO satellite have restricted most of the observations to a range 
of solar elongation between $60^{\circ}$ and $120^{\circ}$. 
This range in elongation is very 
similar to that of DIRBE 
($64^{\circ} - 124^{\circ}$) so both instruments sampled the same 
part of the solar system. 
The zodiacal contribution is based on models by \citet{kel98} and
\citet{rea00} in which the absolute brightness of the zodiacal light 
is normalized to the time variation of the sky brightness observed 
by COBE-DIRBE. 
At each position the DIRBE zodiacal light model takes into account 
emission from the three distinct dust components that constitute the
model: a dust cloud that encompass most of the solar system, a dust ring
around the sun at 1 AU, and dust bands from the asteroid families that extend
from the asteroid belt to the sun. The thermal emission calculated 
by integrating the model along the line of sight is color-corrected 
before it is scaled to the resulting brightness. 

The contribution to the background brightness from Galactic 
cirrus emission
is based on a FIR map developed by \citet{sch98}. 
This map takes advantage 
of both the accurate calibration of the DIRBE FIR data and the higher 
angular resolution of the IRAS 100 $\mu$m map.
The distribution of interstellar dust is highly irregular 
and patchy and there is no available analytical model for its
emission. Consequently, we estimate the emission from 
interstellar dust at all ISOPHOT bands by 
scaling to the 100 $\mu$m map of \citet{sch98},  
using a generic spectrum of the interstellar medium.
The sky brightness is dominated by Galactic emission at all
wavelengths measured with the C200 camera.
The C100 60 and 100 $\mu$m
data also have significant contributions from the interstellar cirrus 
emission, depending on the galactic latitude of the target galaxy. 

We have derived and removed 
the background brightness at the celestial coordinates  
of each galaxy in our sample at the same epoch that the ISOPHOT 
measurements were taken. 
The PHT 32 oversampled
maps were large enough to include a large number of "background" pixels
that allowed us 
to properly evaluate the local sky brightness. The mini-maps have 
more limited 
spatial coverage, but the background flux was estimated using a standard
method based on the determination of fluxes
for each individual pixel as described in \citep{kla00}.
The single pointing measurements (PHT 37/39 and PHT 22) provide us with
a well sampled background field recorded during the off-source pointing.
All ISOPHOT background measurements were then color corrected according to
a representative ``average'' spectral energy distribution (SED) of 
the two-component sky background model described above.

Figure 2 shows the color-corrected ISO backgrounds plotted against 
the  color-corrected COBE-DIRBE measurements toward each galaxy 
in our sample. 
The mean ratios ISO/DIRBE for the color-corrected
background responses are listed for each filter in Table 3.
A very good correlation is seen
between ISO and DIRBE backgrounds at all wavelengths; 
the scatter in the ISO/DIRBE ratio is within
the uncertainty expected in the ISO absolute flux calibrations,
and the dispersion of the data points is reasonably well distributed around
the 1:1 line. 
The single-pointing observations (PHT 22 and PHT 37/39)
show very little dispersion in the 
ISO/DIRBE ratios with average values confined within 
6-7 percent for both the C100 and C200 cameras. 
The oversampled map observations 
show a slightly larger dispersion in the background ratios,
particularly for measurements taken at shorter wavelengths 
with the C100 detector. As seen in Figure
2 the PHT 32 data at 60, 90 and 100 $\mu$m 
show systematic offsets from the 1:1 line by about 20 percent. 
In order to bring the ISO flux scale in line with the COBE-DIRBE flux scale,
we divided each measurement by the appropriate
factor shown in Table 3 prior to any further analysis. 
By this means we have 
normalized our measurements to the DIRBE flux scale. 
This also achieves a cross-calibration between 
the C100 and C200 detectors, 
since COBE-DIRBE has a spectral coverage that extends 
from 25 to 240 $\mu$m. The adoption of the DIRBE flux scale also provides a 
common reference to calibrate the background response 
across all observing modes.

To further verify the absolute calibration of the ISO flux scale, we 
must show 
that the COBE-DIRBE normalization also results in accurate integrated fluxes
for discrete sources. 
This was accomplished by comparing ISO and IRAS photometry 
for a number of elliptical galaxies in the two filters they have in
common, 60 and 100 $\mu$m. 
Although some ISO galaxies were resolved by ISOPHOT, we can safely
consider them compact discrete sources since their extent is only slightly
larger than the FWHM of 
the point-spread function of the two ISOPHOT array 
cameras.  Calibrated survey scans
from the IRAS satellite were extracted using the IRAS Scan Processing and
Integration (SCANPI) software from IPAC.
The software performs scan averaging of the IRAS raw survey data, 
applying several weighting methods for each coadded scan.
The sensitivity gain, which depends on the local noise and number of scans
crossing the target position, is a 
factor of 2-5 above the IRAS Point Source Catalog.
In Figure 3 the integrated fluxes from 30 galaxies in the ISO 
sample (measured using the methods described in Section 6) are 
compared with IRAS fluxes at  60 and 100 $\mu$m, 
revealing a good linear correlation between integrated flux densities.

\section {Spatially Integrated Photometry}

One goal of this work is to determine the integrated flux of 
each early type galaxy in the ISO sample and to improve upon 
earlier surveys of IRAS data.
Both the increased ISOPHOT sensitivity
and the higher spatial resolution can be exploited to improve the
FIR fluxes. 
The method required to extract the integrated photometry 
depends on the ISO 
observing mode and the way the data have been
acquired. About half of the ISO sample has been recorded in raster scans,
either with PHT 32 oversampled maps or with the PHT 22 mini-maps
mode.

The integrated photometry, listed in Tables 4, 5, 6, and 7, refers
to flux densities that have not been color-corrected for 
background emission. 
When a color-correction is mentioned in the text, it refers to
the correction applied to the fluxes under the assumption
that the SED approximate a modified Planck curve ($\beta$=2) 
with temperature 25 K.

{\bf PHT 37/39 \& PHT 22 single-pointing} - These observations 
collect emission from a single pointing in the sky through 
the $2.3^ \prime \times 2.3^ \prime$ and $3^ \prime \times 3^ \prime$
fields of view of the C100 and C200 cameras, respectively.
The photometry was extracted using standard PIA procedures.
Careful inspection of the C100 camera 
pixels was performed before adding the contribution of
each individual pixel, and, when appropriate, noisy pixels were discarded 
and not considered in deriving the source flux.
Table 4 shows the photometric results
along with the statistical error and
background value for each of the galaxies observed in the PHT 37/39
mode.
Table 5 lists the PHT 22 single-pointing photometry along with the errors
and background values. A 3$\sigma$ value represents the upper limit 
to the flux of undetected galaxies.

{\bf PHT 22 mini-maps} - We extracted 
integrated photometry for the eight 
galaxies observed in the mini-map mode using the algorithm
incorporated into the PIA OLP V10 \citep{ric02}. 
The OLP V10 algorithms are appropriate for raster 
observations with a step size equal to the size of a 
pixel since the point-spread function
corrections are defined for pixel separations 
in the two PHOT-C arrays.
As expected, the photometric accuracy
depends on the contrast between the source and the background flux.
In general the mini-maps achieve considerably deeper
photometry than IRAS with an absolute photometric accuracy
better then 30 percent.
Sources observed with the C200 array having fluxes less than about
0.1 Jy are not reliable.
The derived fluxes are 
strictly accurate only for point sources, 
as required by the data fitting algorithm.
If the source is extended, the derived flux 
could be  underestimated by $\sim$15\%, 
depending on its extension.

Five galaxies in the sample are not detected. 
However, even if the undetected galaxies 
are not point sources, we do not expect that 
the error in the upper limits derived with the OLP method 
is significant. For the three objects that show clear emission 
in the central position of the map, we carefully inspected
the background pixels, and found 
no evidence for extended source emission; the
photometry was then performed
using the algorithm in PIA OLP V10. The background
was evaluated as a weighted mean of all the pixels observed
at various positions around the center of the map
with oversampling rates of 6 and 4.
Table 6 shows the flux densities and background flux
values for the eight galaxies observed in this mode;
column 3 and 5 show the flux densities and backgrounds derived
using the PIA algorithm.
The systematic errors, of the order of 30\%,
are larger than the statistical errors and are not included
in the errors reported in Table 6.

{\bf PHT 32 maps} - The integrated flux density 
can be determined from PHT 32 data by either integrating the maps after the 
background is subtracted or by integrating a model that 
best fits the data. 
Both methods have advantages and disadvantages. 
Integrations of 
the raw background-subtracted maps 
are very sensitive to the noise in the data
and may severely affect the integrated photometry of 
low surface brightness objects. 
Furthermore, if any emission
from the source is not covered during the scan, the derived
photometry will be underestimated. 
However, it is very unlikely that incomplete scanning 
affects our photometry since the scans were large enough to sample
the galaxies as seen in visual light down to about 25 mag arcsec$^{-2}$.
Also, most spatially resolved detections with ISOPHOT 
are still quite compact in the infrared, extending only 
slightly beyond the FWHM of the instrumental PSF.

The model-fitting method can recover missing emission
outside the maps, but it depends critically on finding 
parameters that properly describe the data. 
Such modeling also 
depends on the noise level in the maps and 
the spatial sampling on the sky. As mentioned in \S4.1,
the sampling in the spacecraft Z-coordinate is lower than
that in the Y-coordinate, making the determination of reliable
parameters more uncertain. 
In Appendix we describe in detail 
the method used to extract the photometry with model fits.

Since it is unclear which method would be more accurate,  
we decided to derive integrated flux densities 
using both methods.
The derived flux densities are listed in Table 7 
where it is seen that there are no significant 
discrepancies between the direct integration and  
model-fitting methods.
We adopted an empirical approach in evaluating the noise in 
the PHT 32 maps. 
Aperture photometry was performed on 
a number of independent areas spread over the map, avoiding any 
signal from the source. Our raster maps are always large
enough to provide a significant number of pixels free from
source emission. 
The measured scatter in the derived signal was used 
to estimate the noise in each map. 
For non-detections we derived upper limits to the flux density corresponding
to the 3$\sigma$ noise level. 

{\bf CAM 01 maps} - We performed aperture photometry on the 
sky-subtracted frames. At 15 $\mu$m the galaxies are 
resolved by ISOCAM, and some appear very extended.
Since two galaxies have been observed with a smaller 
pixel scale of 3$^{\prime \prime}$, 
we used  an aperture of $60^{\prime \prime}$ in radius 
to be consistent through all our sample. 
Table 8 shows the photometry in the mid-IR filters for the
galaxies detected by ISOPHOT in a least one band.

\section {Results}

The ISOPHOT far-IR fluxes are presented in Tables 4, 5, 6 and 7.
The far infrared emission from early-type galaxies
is not large. Typical fluxes are in the range 0.1- 0.3 Jy at 60 $\mu$m
and 0.3 - 1.5 Jy at wavelengths up to 200 $\mu$m. 
While FIR non-thermal emission from AGN in these galaxies 
can never be completely ruled out, 
many of the elliptical galaxies detected with ISO are rather 
radio-quiet and have FIR spectra that show 
no evidence of non-thermal source contamination.
Nevertheless, when examined in detail many sample galaxies 
have peculiarities of some sort: an anomalously large amount
of cold gas, indication of a recent merger or interaction with
a companion, a counterrotating core, shells, etc. (see Table 2).

Of the 54 galaxies in our ISO sample (we exclude
the four dwarf galaxies NGC 147, NGC 185, NGC 221,
and NGC 2328 from this analysis), only 28 are 
detected in at least one ISO band at the
3$\sigma$ level, a detection rate of $\sim 50$ percent. 
However, given the incompleteness of our sample, 
this detection rate may not be 
representative of early-type galaxies in general. 
In optical studies the dust detection rate for early-type galaxies
is generally larger 
($\sim 80$ percent, \citet{van95}; \citet{fer99}).
For the 39 elliptical galaxies in our sample, only 16 were detected
by ISO ($\sim 41$ percent), while for S0 and later types 
11 out of 14 were detected ($\sim 79$ percent).
The detection rate is similar if we remove the known radio galaxies
from the sample.
Our detection rates for E and S0 galaxies are 
consistent with those found by \citet{kna89}
with a larger sample of early-type galaxies 
observed with the {\it IRAS} satellite
($\sim 45$ and $\sim 68$ percent for Es and S0s respectively).
\citet{bre98} re-examined 
a subsample of the \citet{kna89} IRAS  
galaxies in which peculiar galaxies,
AGN and galaxies with possible observational problems were removed. 
Using this more limited subsample, 
they found a much lower detection rate of $\sim 12$ percent. 
Most of the galaxies in their IRAS subsample 
also have rather large amounts of warm and cold gas.

In estimating the FIR emission from interstellar dust 
in elliptical galaxies,
it is often assumed that the dust grains ejected from 
evolving stars 
are rapidly and widely dispersed into the hot interstellar gas
\citep{tem03}. 
The dust then radiates the energy received by
starlight and thermal electrons during the grain sputtering
timescale. 
The predicted FIR emission
from nearby giant E galaxies
produced by this radiation model is just below the limiting flux
observable with ISO.
However, it may be more plausible that 
the dust ejected from the stars remains with the associated 
stellar gas as it is heated, producing a very inhomogeneous 
interstellar dust distribution 
\citep{mat03}.
When the dust is locally more concentrated, 
the cooling of thermal electrons as they inelastically collide  
with dust grains can dominate over the usual 
thermal radiative cooling from the hot gas.
Consequently, the FIR emission from the grains 
also includes the thermal energy of the hot ambient gas. 
If the total production of dusty gas 
by stellar mass loss in a large elliptical galaxy 
is ${\dot M}$, the maximum expected infrared emission is at least 
$L_{FIR} \sim 5 {\dot M} k T / 2 \mu m_p \sim 
2 \times 10^{41} (T / 10^7~{\rm K})
({\dot M}/M_{\odot}~{\rm yr}^{-1})$ erg s$^{-1}$, where $T$ is 
the (maximum) 
temperature of the hot dusty gas, $\mu \sim 0.62$ is the
molecular weight and $m_p$ is the proton mass.
For typical total mass loss rates from galactic stars, 
${\dot M} \sim 1$ $M_{\odot}$ yr$^{-1}$, 
this is comparable to the FIR luminosities observed by ISO. 
Since ${\dot M} \propto M \propto L_B$ and 
$T \propto \sigma_*^2 \propto L_B^{1/2} \propto M^{1/2}$, 
we would expect $L_{FIR} \propto L_B^{3/2}$ on this 
emission model from galaxies with approximatly constant 
stellar $M/L_B$ and mean stellar age.

It has also been proposed 
that the majority of the ellipticals have acquired
their dust by an accretion event (e.g. \citet{for91}).
This possibility is supported by a comparison of the 
FIR and optical luminosities of the detected galaxies. 
In Figure 4 we plot $L_B$ versus the FIR luminosity
in the various ISO bands
$L_{\lambda} = F_\lambda \Delta_\lambda 4 \pi D^2$, 
where $F_\lambda$ is
the observed flux density at wavelength $\lambda$, $\Delta_\lambda$
is the bandwidth of the filter and $D$ is the
source distance.
Since the emission in the various
bands may be produced by dust with different origin, distribution,
and temperature, we compared $L_B$ with the luminosities in each FIR band,
rather than use the total $L_{FIR}$.
While one dust component might correlate with $L_B$, another may not.
This, in principle, could reveal correlations between $L_B$
and the emission of each dust component that could have been hidden if
the integrated FIR luminosity was used.   
In Figure 4 we combine galaxies 
detected at $90$ and $100$ $\mu$m (panel b) 
and galaxies detected at 150, 170 and
180 $\mu$m (panel c), 
because of the large bandpass overlap of these filters.
Elliptical galaxies are indicated with filled circles, S0 galaxies
and later types are shown with open circles.
No obvious correlation between $L_B$ and any
of the ISO band luminosities is evident (we verified that
fluxes are also uncorrelated, as expected). 
This strongly suggests that the radiating dust is not 
associated with stellar mass loss but is more likely 
to have an external origin.
On the other hand, \citet{bre98} found a correlation
$L_{\rm FIR} \propto L_B^{1.61}$ in their subsample
of 15 elliptical galaxies 
with IRAS detections, which is similar to our 
predicted variation for internally produced dust. 
They avoided peculiar galaxies, which may
also introduce a random disturbance in the $L_{\rm FIR} - L_B$
relation.

The two galaxies in the uppermost left region of Figure 4,
with very high $L_{\rm FIR}$ and relatively low $L_B$,
are NGC 3928 and NGC 5666. These are very peculiar objects.
NGC 3928 has an uncertain classification. Optically, it shows
a small spiral feature embedded in an elliptical envelope.
Its unusual blue
colors and diffuse radio emission indicate starburst activity
in the central region, possibly associated to the capture
of a small dwarf galaxy by an otherwise normal elliptical galaxy
\citep{tan87}. It has a large amount of
cold gas ($\gta 5 \times 10^8$ M$_\odot$).
NGC 5666 has also an uncertain morphology. The large FIR
luminosity may come from dust associated with the very extended
and massive HI envelope \citep{lak87}.

It is also interesting to compare the FIR and mid-infrared 
emission at $\sim 15$ $\mu$m.
The mid-IR surface brightness is often distributed like the
galactic stars (\citet{kna92}; \citet{ath02}), 
which has led to the identification of mid-IR with
warm circumstellar dust 
recently ejected from evolving stars. 
Analyses of the integrated IRAS luminosity at 12 $\mu$m 
reveal that $L_{12\mu{\rm m}}$ is 
roughly proportional to $L_B$ (\citet{jur87}; \citet{esk95}), 
although with significant scatter. In general, the mid-IR emission
may have two components, one due to circumstellar dust which should 
be proportional to $L_B$, and one due
to additional interstellar dust, possibly accreted. 

To gain insight concerning the origin of the mid-IR, 
in Figure 5 we plot the ``mid-IR excess'' $L_{15\mu\rm{m}} / L_B$
against $L_{FIR} / L_B$ for the galaxies detected at both
15 $\mu$m and at least one of the ISO FIR bands. 
ISO data for $L_{15\mu\rm{m}}$
from \citet{fer02} have been added to the 
12 $\mu$m or 15 $\mu$m data
described in section 4.5.
The $L_{15\mu\rm{m}} / L_B$ ratio is not constant among the galaxies,
implying varying amounts of warm dust in some galaxies. 
While only a weak correlation
of $L_{15\mu\rm{m}} / L_B$ with $L_{FIR} / L_B$ is apparent
among the elliptical galaxies, such a correlation becomes
tighter for the S0 galaxies,
pointing toward a common origin for the
infrared enhancement in the mid and far infrared.
Ellipticals seem to continue the
correlation defined by the S0 galaxies at the faintest end
of the IR emission, albeit with more scatter.

In Figure 6 we compare the FIR and X-ray luminosities 
for the subsample of ISO-detected galaxies with $L_x$
listed in \citet{osu01}.
Upper limits for $L_x$
are indicated with left-pointing triangles. 
If galaxies with very low $L_x$ have less hot interstellar gas 
because of galactic outflows (e.g. \citet{bro98}), we might
expect an anticorrelation between $L_{\rm FIR}$ and $L_x$,
since grain sputtering by hot gas significantly
reduces the dust emission (Tsai \& Mathews 1995, 1996).
However, no clear correlation 
is seen in our data. 
In the galactic merger hypothesis, 
the vertical scatter in Figure 6 
might be interpreted as another manifestation
of the random character of the dust accretion process. 
For galaxies in our sample with low $L_x$, 
a large fraction of the X-ray emission has a stellar origin.  
With this in mind, we also verified that 
$L_{\rm FIR}$ does not correlate with the {\it gas} X-ray luminosity,
defined as the difference between the observed $L_x$ and the 
collective X-ray luminosity from X-ray binaries 
assumed to be
$L_{\rm x, discr} = 6 \times 10^{29} (L_B/L_{B,\odot})$ erg s$^{-1}$
(e.g. \citet{sar01}).

We note that many galaxies with measured ISO
emission show peculiarities indicative of 
galactic interactions or mergers. 
About half of the elliptical galaxies detected in the FIR 
also have large masses of cold gas (cf. Table 2), a rare property of
ellipticals as a class. For several other ellipticals no
data on the presence of cold ISM are available or the upper limits
are not stringent. According to the usual interpretation, the cold
gas has an external origin (e.g. \citet{wik95}; \citet{huc95}).
 On average we find that galaxies
that are luminous in the FIR also 
tend to have a large amount of cold gas,
in agreement with \citet{bre98}.

\subsection{Remarks on Galaxies with Companions}

A few galaxies in the sample have a companion at small projected
distance. For these objects the FIR flux may be contaminated by
the presence of the companion. Below we briefly discuss the most
relevant cases.

{\bf NGC 2293} - This galaxy forms a pair with its interacting
companion NGC 2292. The two galaxies, separated by about
0.8 arcmin, have very similar optical luminosities and morphologies. 
The ISOPHOT map is large enough to encompass
the emission from the two galaxies, and 
the brightness distribution at 200 $\mu$m is very extended 
and elongated in the axial direction connecting the two galaxies.
Since the spatial resolution is not high enough to resolve
each component, we suspect that the derived flux density
takes contribution from both galaxies.
The pair has not been detected at 60 and 100 $\mu$m.

{\bf NGC 4105} - This galaxy is strongly interacting with
its nearby neighbor NGC 4106.
Their angular separation $1.2^{\prime}$ corresponds to a projected
linear separation of only 12 kpc at a distance of 35 Mpc.
The massive tidal arms in NGC 4106, very prominent in visible light,
have evidently been pulled out by the
encounter although no distortion of the
elliptical isophotes is seen in NGC 4105 \citep{san94}.
The pointing for the
NGC 4105 ISO observation was accurately chosen to minimize the
contribution to the flux from its companion.
An overlap of the ISOPHOT-C
field of view with an optical image of the target shows that
the optical image of
NGC 4106 lies outside the recorded field.
Although we cannot accurately quantify the contribution
of NGC 4106 to the FIR flux from NGC 4105,
we do not expect it to be very significant.

{\bf NGC 4649} - This galaxy is not detected in 
the far-infrared at any band, but its spiral companion, NGC 4647,
is prominent at all three wavelengths.  The angular resolution
of the oversampled map nicely resolves NGC 4647. Located about
$2^{\prime}$ North-East of NGC 4649, this spiral companion is responsible
for the high flux detection, 
which was erroneously attributed to the large elliptical 
in previous discussions of the IRAS observations. Our 
analysis of the optical and FIR images, shows that there is no 
FIR flux contribution from NGC 4649 (Temi et al. 2003).

{\bf NGC 5353} - Photometry for this galaxy
may also be slightly affected by its companion NGC5354, displaced
$1.3^{\prime}$ to the north.
Both are early type galaxies.
Although it is somewhat
difficult to estimate, we do not expect NGC 5354
to contribute very much to the ISO FIR flux observed for NGC 5353
because careful spacecraft pointing placed
NGC5354 almost completely outside the detector beam.

Finally, two galaxies with companions, NGC 4291 and NGC 6876, 
were not detected at
any of the three bands (60, 100, and 180 $\mu$m) observed.

\subsection {Spectral Energy Distribution and Dust Temperature}

The FIR spectral energy distributions (SEDs) 
for the galaxies detected by ISO are shown
in Figure 7. The color-corrected ISOPHOT flux densities 
are shown with $1 \sigma$ errors. 
For all sample galaxies 
the SED peaks at wavelengths greater than 100 $\mu$m, indicating 
the presence of a cold dust component that previous IRAS observations 
could not have detected. 

Dust temperatures and masses can be derived by fitting modified
blackbody functions to the SEDs. However, such fits may not be unique. 
The mass absorption coefficient $k_{\lambda}$,  
the exponent $\beta$ of the dust emissivity, 
$\epsilon_{\lambda} \propto \nu^{\beta} B_{\nu}(T_D)$, 
and the number of dust components used in fitting the 
FIR data are all intervening factors in determining the goodness
of the fit. Unfortunately, both $k_{\lambda}$ and $\beta$  are
quite uncertain, and their values are still a matter of debate.
As reported by \citet{dun01},  \citet{kla01},  and \citet{ben03}, 
the FIR-submillimeter SEDs may be well fitted with 
different combinations of the  emissivity index $\beta$ and the 
number of dust temperature components used in the fit, 
so the interpretation is degenerate. 
Interpretations of the FIR SED are further complicated by 
our poor knowledge of the emissivity index $\beta$. 

Most of our detected galaxies have four data points, probing the
peak of the dust emission and the start of the Rayleigh-Jeans
region. 
NGC 4261 has been observed in two different runs with both the
PHT 32 and PHT 22 observing modes.
A total of six FIR photometric points are available to constrain its
temperature.
While NGC 4261 is a strong radio source, there is no evidence
of nonthermal emission at ISO wavelengths.
For NGC 3557 fluxes are available only
at 60 and 100 $\mu$m.
Four galaxies in the sample have been observed with the C200 array camera
with one filter only; in order to estimate their dust temperature
we acquired photometric points at 60 and 100 $\mu$m using IRAS
sky-survey data, processed with SCANPI.

Although the ISO data offer a great improvement 
compared to  previous 
dust temperature determinations in early-type galaxies, which were 
based on IRAS data only, the SEDs are not
sufficiently sampled to allow us to establish strong and definitive
interpretation of the FIR dust emission if $\beta$ is allowed 
to be a completely free parameter. 
There are no sub-millimeter data available for the galaxies
in our sample. 
Fortunately, theoretical absorption efficiencies for typical 
silicate or carbon grains suggest $\beta \approx 2$ and 
$\beta > 1$ at very long wavelengths 
because of the Kramers-Kronig relations. 
With this in mind we explore 
two possible models for fitting the SEDs: a 
single-temperature blackbody model 
modified with variable $\beta$, and a 
two-temperature modified blackbody model with fixed $\beta = 2$.

\subsubsection {Single modified blackbody}

The modified blackbody emission from dust at a single 
temperature is 

\begin{equation}
F_\nu = A \nu^\beta B_{\nu}(T_D)
\end{equation}

\noindent
While the amplitude $A$, the dust temperature
$T_D$, and $\beta$,
are all free parameters, values of the exponent $\beta$ 
are constrained
to the range $ 1 \leq \beta \leq 2$, guided by representative values
of the dust emissivity in the Milky Way and 
other well-oberved external galaxies.
Fits to equation (1) 
are performed by minimising the $\chi^2$ and evaluated 
using the reduced $\chi^2$ to account for the number of 
data points that sample the SED in each galaxy.
The index $\beta$ affects the slope in the Rayleigh-Jeans
region:  small values produce a flatter slope and large values
tend to make it steeper. 
Since the wavelength coverage in our sample is limited to 
$\lambda \le 200$ $\mu$m, the Rayleigh-Jeans tail is not fully sampled, 
making the $\beta$ fitting procedure
uncertain. 
For example, \citet{kla01} found 
significant discrepances in $\beta$ between fits to 
the full galactic FIR spectrum as opposed 
to using only the FIR data points.

Aware of this limitation, our first single-temperature
fits to the SEDs 
regard $\beta$ as a free parameter.
The fitted parameters are shown in Table 9 together with the
reduced $\chi^2$. 
Most of the $\beta$ values are very small, approaching the lower 
allowed limit of 1. If $\beta$ is unrestricted, the best fit would 
provide unrealistic values well below 1 for the majority of the
galaxies. This is clearly due to the lack of 
sufficient data points in the 
sub-millimeter region, and is a manifestation of an unreliable fit
rather than an indication of unusually low $\beta$. 
Only 8 of the 28 galaxies detected have acceptable 
fits under the $\chi^2$ criterion. The large majority
of the galaxies could not be fitted with a single-temperature
model. 
The mean value for the dust temperature $T_D$ derived by 
the fit is 27.9 K, spanning a range from 22 to 43 K.

It is difficult to reconcile specific estimates
of $\beta$ with a unique dust composition, size distribution and 
temperature (see Dunne \& Eales, 2001, for a short review 
on $\beta$ values in the literature). 
Almost all observations 
of external galaxies report values of $\beta$ between 1.6 and 2, 
with $\beta$=2 be the most accredited (e.g. Chini et al. 1989, Chini
\& Krugel 1993, Bianchi et al. 1998). 
However, FIR emission from grains with small $\beta$ 
may contribute less 
to the SED when averaged over the entire galaxy.
Nevertheless, following previous observations, 
we also performed single modified blackbody fits with
fixed $\beta$=2. 
These temperatures are reported in Table 9 along with the reduced
$\chi^2$. With $\beta$ fixed at 2, 
only four galaxies are compatible with a single-temperature
model and the other 24 require at least two dust components. 
The temperatures have a mean of 23.8 K with a range of 19 - 34 K.
With this steeper $\beta$=2 emissivity, the average
temperature is colder by $\sim$ 4 K, confirming a trend already 
reported in the literature \citep{ben03}.

Errors in the dust temperature arise from uncertainties in the
ISOPHOT photometry and the assumed grain emissivity.
Since our ISOPHOT data are calibrated to the COBE-DIRBE
flux scale, we are confident that calibration errors are lower than
the generally assumed 30 percent except for the faintest objects with
very low signal-to-noise ratios.
We estimated the error in the derived temperatures by varying
the fluxes at the shortest and longest wavelength:
we increased the 60 $\mu$m flux by 20\% and decreased the 200 $\mu$m 
flux by 20\% and vice-versa.
The estimated errors are of the order $\sim$ 2 K for the entire 
sample.

\subsubsection {Two modified blackbodys}

In the previous section we showed that 
the SEDs for most of the sources could not be fitted with a 
single-temperature model if $\beta = 2$.
We now consider spectral models in which the SED is fit 
with two dust components (warm and cold), 
each having modified blackbody spectra given by equation (1), and
assuming a fixed $\beta$=2 dust emissivity index. 
Four free parameters describe the temperatures and amplitudes of the
two modified blackbodies and these parameters were used 
to construct $\chi^2$ fits to the data.
Six galaxies detected using the PHT 37/39 mode have been
observed in three ISOPHOT bands only, 
for which the two-component fits are 
under-constrained. For these galaxies we 
assume that the warm dust temperature
is 43 K, the mean warm dust 
temperature derived for galaxies observed at 
four or more wavelengths. 
We note that the temperature of the cold dust is 
insensitive to relatively large variations of 
the assumed warm dust temperature for these galaxies. 
The fitted warm and cold dust temperatures are listed in Table 9. 
The cold dust component 
temperatures range from 17 to 24 K with an average of 19.5 K. 
In all likelihood, the dust temperatures are not 
confined to two temperatures, but span a continuum 
between them.
Finally, our dust temperatures 
compare well with the temperature $\sim 18$ K found by Stickel
et al. (2003) for M86. 
Using the same method descibed in the previous section,
the estimated errors on the derived temperatures 
due to uncertainties in the flux densities are $\leq 10\%$
for both warm and cold components. 
Neither the warm nor cold dust temperatures correlate with $L_B$.

It is interesting to compare the derived temperatures of the four galaxies,
NGC 1395, NGC 1553, NGC 4261, NGC 4552,
for which the SEDs are consistent with both fitting models.
For all but one (NGC 4552) the one-temperature fit is 
warmer by less then 2 K relative to the cold temperature fit 
in the two-blackbody model. 
This is within the estimated error in the derived
temperature due to uncertainties in the fluxes.
For NGC 4552, the discrepancy in temperature between the 
two fitting model is $\sim$ 5 K.

We also estimated the dust temperature using only 
the 60 and 100 $\mu$m ISO flux densities, the same
wavelengths available to IRAS.
A single component dust temperature fit to the ISOPHOT
60 and 100 $\mu$m flux densities yields a mean dust temperature 
for all detected galaxies of 29.1 K, which is 
significantly larger than the mean cold dust temperature
19.5 K found using all the ISO photometric data.

\subsection {Dust Masses}

The dust masses can be estimated 
using the parameters that fit the dust SED and the standard relation
\citep{hil83,dra90}:

\begin{equation}
M_d = {F_{\lambda} D^2  \over k_{\lambda} B_{\lambda}(T_D) }  
\end{equation}

\noindent
where $F_{\lambda}$ is the observed flux density at wavelength $\lambda$, 
$D$ is the 
galaxy distance, and $k_{\lambda}$ is the dust opacity. 

The derived dust masses depend critically on the
dust temperatures, the dust emissivity index $\beta$, and 
the opacity $k_{\lambda}$. Dust masses are notoriously uncertain 
(e.g. Devereux \& Young 1990). 
We adopted the opacities calculated by Li \& Draine (2001),
81 cm$^2$ g$^{-1}$ and 9 cm$^2$ g$^{-1}$ at 60 $\mu$m and 180 $\mu$m,
respectively, which are representative of currently
considered values.
The masses were derived from the flux values at 180 $\mu$m;
for the two-blackbody model the ``warm" dust component
was estimated using fluxes at 60$\mu$m. The value of $k_{\lambda}$
was adjusted to be consistent with the value of $\beta$ assumed for 
the FIR emission. 
Table 9 lists the dust masses calculated for each of the three
models used in fitting the SEDs.
Higher temperatures derived in the single-blackbody
fits correspond to lower calculated dust masses.
Also, dust masses are higher when calculated with larger values 
of $\beta$. The reason for this is twofold: (i) larger 
$\beta$ results in lower dust temperatures 
that decrease the value of 
$B_{\lambda}(T_D)$ in the denominator of equation (2), 
and (ii) $k_{\lambda}$ 
at 180 $\mu$m is lower when larger $\beta$
values are used. 

Clearly, the FIR emission from dust at a single 
temperature cannot fit the spectra of most of our 
early type galaxies.
However, a combination of a warm and cold dust
components provides an acceptible representation of the 
observed FIR. 
Although the two-temperature fits must be taken with
some caution, we consider the dust masses derived in this way
the best estimate 
of the dust content in these galaxies. 

The mass of dust in the cold component is 
about 3 orders of magnitude larger then the mass of warm dust. 
The dust content is not small. Several galaxies have dust
masses near $10^7 M_\odot $.
As pointed out by \citet{dal02}, the dust masses may
be underestimated by a factor of few when the SEDs are fitted 
with one or two blackbodies insted of considering a 
(more realistic) continuous
range of blackbodies corresponding to a range of dust 
heating evironments. 

The total dust mass does not correlate with $L_B$ 
over the limited range of $L_B$ represented in the sample. 
This again may indicate a stochastic external origin for the dust.
Table 9 also shows upper limits to the dust
mass evaluated for galaxies that were not detected with ISO. These
upper limits have been constructed using the 3$\sigma$ fluxes 
and assuming a mean dust temperature of 19.5 K.
This is not strictly a true upper limit to the dust mass since
the actual temperature may be lower then the assumed one, 
but it provides a useful estimate. 

It is important to stress that photometry at wavelengths
greater than $100 \mu$m are necessary to 
view the dust SED around
its peak emission and are therefore 
critical in determining the total dust mass.
Estimates 
of dust temperatures based on the flux ratio F$_{60}$/F$_{100}$ can
severely underestimate the derived dust mass. 
In Figure 8 we compare our estimated dust masses with values derived
from IRAS temperatures 
(Roberts et al. 1991; Goudfrooij et al. 1994; Bregman et al. 1998) 
scaled to our assumed distances.
The lower dust temperature indicated by ISO 
increases the estimated dust mass by factors of 
$\langle M_{\rm d,ISO}/M_{\rm d,IRAS}\rangle
\sim $12 , 24, and 16 when compared to the IRAS masses
given by \citet{rob91}, \citet{gou94}, and \citet{bre98},
respectively.

\subsection{Angular extent of the dust emission}

The spatial distribution of the dust in elliptical galaxies
is poorly known. Most early-type galaxies show dust absorption
features in the very center (e.g. van Dokkum \& Franx 1995), 
but very little is known about the
presence of an extended component of dust. It is well known that
for elliptical galaxies
the dust mass estimated from FIR fluxes is much larger than the dust
mass derived from optical absorption studies (Goudfrooij \& de Jong 1995).
This discrepancy may be solved if most of the dust is smoothly
distributed in a large volume. 

ISO observations in the PHT 32 AOT mode
may shed light on the extension of the FIR emission.
Many of the galaxies observed in the oversampled map mode have
been selected because their D$_{25}$ diameters 
are much larger 
than the ISO telescope beam. For example, with a sky sampling of 
$15^{\prime \prime} \times 23^{\prime \prime}$ at 60, 90 and 
100 $\mu$m the PHT 32 AOTs approach the Nyquist limit, which 
is $\sim$ 17$^{\prime \prime}$ 
for the ISO telescope at 100 $\mu$m.
These observations have the potential to resolve the galaxies 
and to determine if the far-infrared emission is produced 
by dust distributed
throughout the galaxy or by the central dust clouds found 
in optical surveys. 
In principle, the spatial distribution of the FIR emission 
may also reveal information about the controversial
origin of the dust in the elliptical galaxies.
Dust accreted in a recent merger is expected to have an extended,
asymmetric distribution, while internally produced dust should have
a more compact and regular aspect. However, the faint level of the
FIR emission in early-type galaxies and the relatively poor spatial
resolution of ISOPHOT do not allow a secure study of the asymmetries
in the dust distribution.

To identify galaxies in the PHT 32 AOT sample that 
are more extended than 
the PSF at FIR frequencies, we developed
an image modeling procedure to apply to the oversampled map data. 
We performed a beam-fit for sources that appear point-like 
and modeled sources with a two-dimensional gaussian image 
if there is evidence for some detected extension.
Since there is no analytical description of the shape of
a true beam, we first constructed a beam-model based on a
circular gaussian. 
Using the PIA Map Simulator, we constructed noiseless maps
of a point source by setting the
sampling parameters in the same way they were set in acquiring the
actual PHT 32 data.
After correcting the simulated maps for transients, 
the derived widths
of the gaussian beam-fit model correspond to 
$\sigma$=$18^{\prime \prime}.60$,
$20^{\prime \prime}.25$, $20^{\prime \prime}.40$,
$39^{\prime \prime}.90$, $40^{\prime \prime}.80$,
and $41^{\prime \prime}.10$ at 60, 90, 100, 150, 180, and 200 $\mu$m,
respectively, where the FWHM = 2.354$\sigma$. 
We used these model beams to fit pointlike sources.

The observed emission from 
extended sources was found by convolving
the circular beam-fit model with an elliptical
gaussian source model. 
The parameters that must be adjusted to achieve a fit to 
an extended source 
are the amplitudes, rotation angles, and sigmas of the two
gaussians along the principal axes of the elliptical source model. 
We used an optimization routine from the IDL Library to minimize
a multi-parameter function, determining  the parameters 
for the best fit. 
The PHT 32 maps are undersampled
in the cross-scan ($z$) direction. 
Therefore, before performing each model fit,
we re-sampled the data in the $z$ direction 
using a gridding algorithm in PIA.

This simple image fitting procedure is well 
suited to match the surface brightness profiles 
of elliptical galaxies 
where the galaxy morphology is rather smooth and regular.
Two galaxies in the sample - NGC 2293 and NGC 4649 - are either
in close interaction with companions or have a complex
brightness profile.
As a consequence our elliptical model is not 
appropriate to describe the light profiles for these two galaxies 
so we have not attempted to model them.

Among the 10 galaxies
that have been detected with PHT 32, 
four appear to be resolved by ISOPHOT
at least in some of the bands observed. In some cases the
angular extent derived by the model-fit is only marginally larger
than the instrumental beam, making our estimate rather uncertain.
Table 10 shows the angular extent of the infrared emission 
of the source ($\sigma_{S,major}$, $\sigma_{S,minor}$) at the
various wavelengths
and other parameters of the model fits for 
these galaxies with extended FIR images.
The gaussian scales 
$\sigma_S$ and $\sigma_T$ refer to the 
source model and the the convolution
of the source model with the beam model, respectively, 
so $\sigma_S \le \sigma_T$ is always expected. 

At 200 $\mu$m NGC 5666 
shows a complex and extended surface brightness structure 
for which a multi-component model would be more appropriate 
than our elliptical gaussian model.
Consequently, the angular size for NGC 5666 in Table 10 
based on our gaussian model should be regarded as approximate.
NGC 4261 and NGC 5173 have been clearly detected at the wavelengths 
shown in Table 10, but they are faint and their low S/N ratios
make the determination of the ellipsoid position angle quite 
uncertain. 
The extension of the ISO image for NGC 4261 exceeds 
both the size of the central dusty disk 
$\sim 2^{\prime \prime}$ seen in HST images \citep{jaf93} and the 
dust feature $15^{\prime \prime}$ distant 
from the galactic center \citep{mol87}.
For each of the three galaxies in Table 10 that have been observed 
at more than one ISO band, the size of the source $\sigma_S$
increases steadily 
with increasing wavelength and often exceeds the optical 
half light radius $R_e$. 
Such a behavior would be expected if the dust is heated 
by stellar radiation which decreases from the galactic center

Figure 9 shows the observed brightness profiles (histogram) 
together with their model fits (solid line). 
The beam-fit profiles for point sources are shown with dashed lines.
To illustrate the likelihood or 
degree of spatial extension, we show both the 
model-fit and beam-fit for many galaxies. 
For each FIR band the middle panels in Figure 9 show fits to the
row positioned on the galaxy center along the scan direction (y-axis).  
The top and bottom panels show off-source 
parallel scans taken at 
two symmetric positions on both sides of the source. The offset in 
the z direction (the cross-scan direction) is noted in each plot.

We also compared the effective radius $R_e$ of the galaxies
that were not resolved with PHT 32 with the ISOPHOT beam sizes. 
Among the 6 detected galaxies that appear as pointlike,
two (NGC 3998 and NGC 5813) have effective radius
$\sim$ 2 times larger than the $ \sigma$  $\sim 20^{\prime \prime}$
of the beam at 60 and 100 $\mu$m. 
At least for these two galaxies, this is indicative that 
most of the FIR is coming from the central dust cloud
known to exist in both galaxies.
At larger wavelengths, the effective radius become comparable 
to the size $\sigma$ of the beam. 

\section {Conclusions}

We extracted FIR data 
for a sample of early-type
galaxies from the ISO archives. The data extracted 
are described in Table 1 and the sample is described in Table 2.
It includes 53 giant galaxies: 39 ellipticals 
and 14 S0 (or later) galaxies, as classified in the RSA catalog.
The data reduction and calibration  
give very consistent results, both for point
sources and extended objects. 
We tested and verified 
the photometric performance of ISO 
by comparing with data obtained previously. 
A comparison with IRAS observations of the same early-type 
galaxies serves as an independent 
calibration check for the ISO response to faint discrete sources. 
The integrated fluxes found with ISO  
match IRAS fluxes within a few percent for point sources.
Moreover,
the sky brightness is in good agreement with COBE-DIRBE results.
Based on these comparisons, we are confident that a reliable
ISO calibration has been established.

The most important contribution provided by ISO
are the data at longer wavelengths ($\lambda > 100 $ $\mu$m). This allows
a much more accurate determination of the dust temperature 
and dust mass than was possible with previous IRAS data.
We find that the FIR spectral energy distribution requires
emission from dust with at least two different temperatures. 
The mean temperature of the hot dust component is 43 K,
while the cold dust, which is always the dominant component by mass,
has a mean temperature of 20 K.
By contrast, the dust temperature derived by fitting the IRAS data 
results in a single temperature that is about 10 K higher. 
We suspect that our two-temperature dust models represent the 
extremes of a continuum of dust temperatures.

Typically, the dust mass ranges from $10^5$ to 
more than $10^7$ M$_\odot$.
The dust masses derived from ISO data are on average 
at least ten times 
greater than those previously estimated from IRAS observations
\citep{rob91}. 
Using the Galactic dust to gas mass ratio, $\sim 0.01$, 
the total mass of cold gas associated with this dust could be 
in the range $10^7$ to more than $10^9$ M$_\odot$.

The lack of any recognizable correlation 
between $L_{FIR}$ or dust mass  
and $L_B$ suggests that the dust in 
at least some of our sample galaxies 
has been acquired externally in a merger event. 
While the ISO data provide new and useful information 
about FIR emission from early-type galaxies, 
the sample of archival data is far from ideal.
It is likely that 
many galaxies selected for ISO observations were chosen because 
they were known to have large IRAS fluxes or 
large masses of cold gas.
Elliptical galaxies with cold HI or molecular gas are rare 
and probably unrepresentative. 
Such galaxies are likely to have experienced unusual 
dust-rich mergers that could mask physically 
interesting correlations expected from internally produced dust. 
But it is unclear if mergers are the dominant source of 
dust in E and S0 galaxies.
For example, if most of the FIR is emitted by central dust clouds,
these clouds could be disturbed at irregular intervals by
low level AGN activity in the galactic cores,
creating the stochastic $L_{FIR}$ variation in our sample.
Furthermore, 
the sensitivity of IRAS and ISO are not sufficient to 
detect FIR emission from 
many optically luminous E galaxies such as NGC 4472.
While there is clear evidence of 
circumstellar mid-IR emission from mass-losing red giants, 
the FIR emission studied so far cannot be unambiguously 
attributed to the same dust when it becomes truly interstellar.
To make progress, it will be necessary
to obtain deeper FIR observations
from a more representative, optically-selected sample.

\acknowledgments

The data presented in this paper were taken in
guaranteed observing time made available by
the following PIs: Bregman, J. N., Knapp, G.,
Macchetto, F., Norgaard-Nielsen, H. U., Renzini, A. 
and Vigroux, L.; we thank the referee for a careful 
reading of the manuscript, and for a number of comments 
and suggestions that helped to made it clearer.
This study is supported by NASA grant NRA-01-01-ADP-032
for which we are very grateful.
FB is supported in part by grants MURST-Cofin 00
and ASI-ARS99-74. WGM is supported by 
NASA grants NAG 5-8409 \& NAG 5-13275 and NSF grants
AST-9802994 \& AST-0098351.
Finally, this research has made use of the NASA/ IPAC 
Infrared Science Archive, which is operated by the Jet 
Propulsion Laboratory, California Institute of Technology, under 
contract with the National Aeronautics and Space Administration.

\appendix  {\hspace{2.5in} \textbf {APPENDIX}}

Following the procedure described in detail by Tuffs et al. (2002), 
we developed a scheme to analyze our PHT 32 data to detect 
and quantify the spatial extent of the FIR emission. 
First we determine an appropriate circular 
gaussian beam-fit for pointlike sources 
and then we compare this to an ellipsoidal beam-model for 
the source.
Extended sources are assumed to have gaussian profiles along 
each axis of the ellipsoid. 
An optimization routine from the IDL library is used to minimize
a multi-parameter function and to determine the 
amplitudes, position angle and ellipticity for the best-fit. 
We first constructed a beam-model based on a
circular gaussian as described in  $ \S $ 7.4. To account for
the difference in shape between the true beam profile and the
gaussian 
model, a correction factor must be applied to 
the flux density derived by integrating the beam model. 
We evaluated these corrections by generating a simulated map
of a point source of known flux density using the PIA Map Simulator.
>From the simulated maps corrected for transients,
we found that the derived flux densities were
less than the original value entered in the simulation maps 
at all wavelengths.
The correction factors range from 0.84 at 90 $\mu$m to 0.89
at 200 $\mu$m. We applied these corrections to the computed flux
densities for all beam-fits performed on pointlike sources.
The fit for the extended sources was performed by convolving
a beam-fit model with an  elliptical
gaussian source model; the amplitudes, rotation angles and sigmas of 
the gaussian source model were adjustable parameters in the fit.

\clearpage

\figcaption{Layout of the PHT 22 mini-maps. The top panel shows the
C100 layout for a $3 \times 3$ raster map. Each square represent
a pixel of $46^{\ \prime \prime}$ in size, and a black filled dot marks
each position in the raster grid. The grayscale intensity
changes as a function of the number of observations in that specific position.
The central position has been observed 9 times during the scan, while the four
corners have only one measurement. The thick dashed square represent the
$3 \times 3$ C100 detector array. The bottom panel gives the $4 \times 2$
raster layout for the C200 $2 \times 2$ pixel array; each pixel is
$92^{\prime \prime}$ in size. The central map position
has been observed 4 times. For both arrays the raster step is equal
to the size of one pixel. \label{fig1}}

\figcaption{ISOPHOT backgrounds at several far-infrared filters against
brightness derived from a model based on COBE-DIRBE data.
Data for the background model were color-corrected and evaluated toward
the position of each galaxy and at the same epoch
when the ISO observations were
recorded. Each panel shows measurements at all the available ISOPHOT bands
for a specific observing mode.
Panels (a) through (d) present data for the  PHT 32, PHT 37/39,
PHT 22 mini-maps, and PHT 22 single-pointing, respectively.
Each symbol identifies a wavelength band: ($\ast$) is 60$\mu$m band,
(+) is 90 $\mu$m band, ($\Diamond$) is 100$\mu$m band,
($\triangle$) is 150$\mu$m band, ($\Box$) is 180$\mu$m band,
and ($\times$) is 200$\mu$m band. \label{fig2}}

\figcaption{Integrated flux density of a sample of early-type galaxies
observed with ISO C60 (left panel) and C100 (right panel) filters versus the flux
density measured by IRAS in the 60 and 100$\mu$m bands.
The color-corrected ISO fluxes, have been recorded using the PHT 32, PHT22,
and PHT37/39 observing modes.
The straight line
represents the one-to-one correlation between the ISO and IRAS integrated
fluxes. \label{fig3}}

\figcaption{ISO FIR luminosities in the various bands vs. $L_B.$
In panel (b) we have grouped together galaxies detected at 90 $\mu$m
with those detcted at 100 $\mu$m. Panel (c) includes data
at 150, 170 and 180 $\mu$m. Filled circles indicate elliptical
galaxies, open circles indicate S0 or later galaxies. The four
dwarf ellipticals and NGC 1275 are not shown in this plot.
\label{fig4}}

\figcaption{$L_{15\mu\rm{m}} / L_B$ vs. $L_{FIR} / L_B$. As in Figure 4
panel (b) include data at 90 and 100 $\mu$m, while panel (c)
include data at 150, 170 and 180 $\mu$m. Filled circles denote ellipticals,
open circles S0 or later galaxies.
\label{fig5}}

\figcaption{ISO FIR luminosities vs. $L_x$.
Panel (b) include data at 90 and 100 $\mu$m, panel (c)
include data at 150, 170 and 180 $\mu$m. Filled symbols denote ellipticals,
open symbols S0 or later galaxies. Left-pointing triangles indicate
upper limits in $L_x$.
\label{fig6}}

\figcaption{FIR Spectral Energy Distribution of the detected galaxies in
the sample. Color-corrected flux densities are plotted with their
associated error bars.
The dotted lines show the two modified blackbody functions that
best fits the data. The sum of the "warm" and "cold" fitting functions
are represented by the solid line. The fitting temperatures for the cold
dust component are marked in the upper right corner. For the galaxies
with only three photometric points, a warm dust temperature T=43 has
been assumed. \label{fig7}}

\figcaption{ISO Dust masses vs. IRAS dust masses. Filled circles refer to
IRAS data from Roberts et al. 1991; open squares are from
Goudfrooij et al 1994; open circles are from Bregman et al 1998.
IRAS masses have been scaled to our assumed distances. The solid line
represents the 1:1 relation.  \label{fig8}}

\figcaption{Three surface brightness profiles are shown for each
observed galaxy. The central panel shows the variation along
the scan direction (y-axis) passing through the center
of the source. The upper and lower panels
show similar nearby parallel scans displaced symmetrically
from the source in a perpendicular
direction (z-axis) by the amount shown.
The beam-fits corresponding to pointlike sources
are plotted as dashed lines. The solid lines show
beam-models that are broader than the pointlike response
for those relatively few sources that are extended. }

\clearpage
\begin{deluxetable}{lrrcllccr}
\tablenum{1}
\tabletypesize{\scriptsize}
\tablewidth{0pt}
\tablecaption{Logs of the Observations}
\tablecolumns{8}
\tablehead{
\colhead{Name}                          &
\colhead{$\alpha(2000)$}                &
\colhead{$\delta(2000)$}                &
\colhead{TDT\tablenotemark{a}}          &
\colhead{AOT\tablenotemark{b}}          &
\colhead{Filter}                        &
\colhead{$N\times M$\tablenotemark{c}}  &
\colhead{f.o.v.}                        &
\colhead{$T_{int}$\tablenotemark{d}}    \\
\colhead{}                    &
\colhead{}                    &
\colhead{}                    &
\colhead{}                    &
\colhead{}                    &
\colhead{}                    &
\colhead{}                    &
\colhead{$^{\prime \prime} \times ^{ \prime \prime}$}  &
\colhead{(s)}                 
}

\startdata

NGC 0147 ..........   & 00 33 12.12 &+48 30 20.5 & 80104478 & PHT22      & C60   & $3\times 3$ & 227x227  & 420   \\
                      & 00 33 12.12 &+48 30 20.5 & 80104478 & PHT22      & C90   & $3\times 3$ & 227x227  & 420   \\
                      & 00 33 12.12 &+48 30 20.5 & 80104479 & PHT22      & C160  & $4\times 2$ & 457x273  & 19    \\
                      & 00 33 12.12 &+48 30 20.5 & 80104479 & PHT22      & C200  & $4\times 2$ & 457x273  & 195   \\
\\
NGC 0185 ..........   & 00 38 56.90 &+48 20 10.9 & 78500680 & PHT22      & C60   & $3\times 3$ & 227x227  & 420   \\
                      & 00 38 56.90 &+48 20 10.9 & 78500680 & PHT22      & C90   & $3\times 3$ & 227x227  & 420   \\
                      & 00 38 56.90 &+48 20 10.9 & 78500681 & PHT22      & C160  & $4\times 2$ & 457x273  & 195   \\
\enddata
\tablecomments{Units of right ascention are hours, minutes, and seconds;
units of declination are degrees, arcminutes, and arcseconds.}
\tablenotetext{a}{Target Dedicated Time identifier (TDT)}
\tablenotetext{b}{Astronomical Observing Template (AOT)}
\tablenotetext{c}{Raster points in Y and Z spacecraft coordinates}
\tablenotetext{d}{On-source integration time}
\end{deluxetable}

\clearpage

\begin{landscape}

\begin{deluxetable}{lrrrrrrrrl}
\tablenum{2}
\tabletypesize{\scriptsize}
\tablewidth{0pt}
\tablecaption{Basic Properties of the Sample}
\tablecolumns{10}
\tablehead{
\colhead{Name}                          &
\colhead{D\tablenotemark{a}}            &
\colhead{Log $L_{B}$\tablenotemark{a}}  &
\colhead{Log $L_{X}$\tablenotemark{b}}  &
\colhead{$F_{6cm\tablenotemark{c}}$}    &
\colhead{ T\tablenotemark{d}}           &
\colhead{ type\tablenotemark{e}}        &
\colhead{Log $M_{\rm{HI}}$\tablenotemark{f}}             &
\colhead{Log $M_{\rm{H_2}}$\tablenotemark{m}}            &
\colhead{Notes\tablenotemark{r}}     \\
\colhead{}                    &
\colhead{(Mpc)}               &
\colhead{($L_{B\odot}$)}      &
\colhead{($erg s^{-1}$)}      &
\colhead{(mJy)}               &
\colhead{}                    &
\colhead{}                    &
\colhead{($M_\odot$)}         &
\colhead{($M_\odot$)}         &
\colhead{}
}

\startdata

IC 1459 ..............   & 18.88  & 10.37  & 40.71       &1016.0 &-4.7  & E1       & $\leq $8.64    & \nodata   & d, e$H\alpha$, R   \\ 
NGC 0147 ..........  & 0.65  &  7.92   &$\leq $37.45 &12.0    &-4.8  & dE5         & $\leq $6.21    & \nodata   &          \\
NGC 0185 ..........  & 0.62  &  8.07   &$\leq $37.36 &\nodata &-4.8  & dE3pec      & 5.10           & 5.04      &          \\
NGC 0221 ..........  & 0.72  &  8.36   & 37.77       &\nodata &-4.7  & E2          &  $\leq $6.26   & \nodata   &          \\
NGC 0596 ..........  & 22.28 & 10.21   &$\leq $39.60 &\nodata &-4.3  & E0          &  8.10\tablenotemark{g}   & \nodata  &      \\
NGC 0720 ..........  & 20.80 & 10.38   & 40.61       &\nodata &-4.8  & E5          &  $\leq $8.44   & $\leq $7.77\tablenotemark{n}  &       \\
NGC 0807 ..........   & 64.64 & 10.39   &   -        &\nodata  &-4.8  & E           &  9.92\tablenotemark{h}   & 9.16\tablenotemark{o}  &     \\
NGC 0821 ..........   & 20.99  & 10.16  &$\leq $40.33 &\nodata &-4.8  & E6         & $\leq $7.96     & \nodata  &        \\
\enddata

\tablecomments{
Luminosities and distances are calculated with $H_0=75$ km s$^{-1}$ Mpc$^{-1}$; 
$(^a )$  {data are from the LEDA catalog}; 
$(^b )$  {data are from O'Sullivan et al, 1999}; 
$(^c )$ {``pseudo-6 cm'' luminosity from Roberts et al. (1991)};
$(^d )$ {morphological type code from the LEDA catalog};
$(^e )$ {morphological type from the RSA catalog};
$(^f )$ {HI data from Roberts et al (1991), unless otherwise indicated};
$(^g )$ {from Huchtmeier (1994)};
$(^h )$ {from Huchtmeier, Sage \& Henkel (1995)};
$(^i )$ {from Sadler et al. (2000)};
$(^l )$ {from Knapp \& Raimond (1984)};
$(^m )$ {H2 data from Roberts et al (1991), unless otherwise indicated};
$(^n )$ {from Braine, Henkel \& Wiklind (1997)};
$(^o )$ {from Wiklind, Combes \& Henkel (1995)};
$(^p )$ {from Lees et al. (1991)};
$(^q )$ {from Sofue \& Wakamatsu (1993)};
$(^r )$ {d=dust detected through optical extintion; $H\alpha$=central emission line gas;
e$H\alpha$=extended emission lines gas; R=radio source,M=possible recent merger remnant}
}

\end{deluxetable}

\end{landscape}

\clearpage

\begin{deluxetable}{lrccccccccccc}
\tablenum{3}
\tabletypesize{\scriptsize}
\tablewidth{0pt}
\tablecaption{ISO and COBE-DIRBE response to the background}
\tablecolumns{6}
\tablehead{
\colhead{}    &
\colhead{}    &
\multicolumn{2}{c}{PHT 32}    &
\colhead{}    &
\multicolumn{2}{c}{PHT 37/39} &
\colhead{}    &
\multicolumn{2}{c}{PHT 22 mini-maps} & 
\colhead{}    &
\multicolumn{2}{c}{PHT 22 single-pointing}   \\
\cline{3-4} \cline{6-7} \cline{9-10} \cline{12-13}  \\

\colhead{Filter}              &
\colhead{$\lambda$}           &
\colhead{$_{ISO/DIRBE}$}     &
\colhead{$\sigma$}            &
\colhead{}                    &
\colhead{$_{ISO/DIRBE}$}     &
\colhead{$\sigma$}            &
\colhead{}                    &
\colhead{$_{ISO/DIRBE}$}     &
\colhead{$\sigma$}            &
\colhead{}                    &
\colhead{$_{ISO/DIRBE}$}     &
\colhead{$\sigma$}    \\
\colhead{}                    &
\colhead{($\mu$m)}            &
\colhead{}                    &
\colhead{}          &
\colhead{}                    &
\colhead{}                    &
\colhead{}          &
\colhead{}                    &
\colhead{}                    &
\colhead{}          &
\colhead{}                    &
\colhead{}                    &
\colhead{}
}

\startdata

C60     & 60    & 1.267  & 0.187 &  & 1.030  & 0.144 &  & 1.022 & 0.042   & & 1.038  & 0.087\\
C90     & 90    & 1.220  & 0.097 &  &\nodata &\nodata&  & 0.942 & 0.055   & & 1.056  & 0.094\\
C100    &100    & 1.229  & 0.167 &  & 0.990  & 0.135 &  &\nodata&\nodata  & &\nodata &\nodata\\
C135    &150    & 1.069  & 0.204 &  &\nodata &\nodata&  &\nodata&\nodata  & &\nodata &\nodata\\
C160    &170    & \nodata&\nodata&  &\nodata &\nodata&  & 1.003 & 0.144   & & 1.049  & 0.077\\
C180    &180    & 1.040  & 0.069 &  & 0.934  & 0.138 &  &\nodata&\nodata  & &\nodata &\nodata\\
C200    &200    & 0.932  & 0.230 &  &\nodata &\nodata&  & 0.832 & 0.190   & & 0.948  & 0.167\\
\enddata

\tablecomments{ISO/DIRBE is the mean ratio of the color-corrected background responses
for the two instruments;
$\sigma$ is the standard deviation of the ratios after the ISO backgrounds have been corrected.}

\end{deluxetable}

\clearpage

\begin{deluxetable}{crrrccl}
\tablenum{4}
\tabletypesize{\scriptsize}
\tablewidth{0pt}
\tablecaption{ISOPHOT P37/39 photometry. \label{tbl-2}}
\tablecolumns{6}
\tablehead{
\colhead{Name}                        & \colhead{${\rm \lambda}$}         &
\colhead{ Flux}                       & \colhead{$\sigma_{Flux}  $}       &
\colhead{ F$_{back}$}                 & \colhead{$\sigma_{background}$}&
\colhead{Notes}     \\
\colhead{}                  & \colhead{($\mu$m)}          &
\colhead{(Jy)}             & \colhead{(Jy)}               &
\colhead{($MJy sr^{-1}$)}  &\colhead{($MJy sr^{-1}$)}     &
\colhead{}
}

\startdata

NGC 1332 ....   &  60  &  0.49     & 0.07  & 8.89  & 0.72  &              \\
                & 100  &  1.52     & 0.04  & 6.36  & 0.32  &              \\
                & 180  &  1.56     & 0.06  & 5.45  & 0.22  &          \\
\\
NGC 1553 ....   &  60  &  0.74     & 0.07  & 6.57  & 0.50  &              \\
                & 100  &  1.01     & 0.04  & 4.49  & 0.14  &              \\
                & 180  &  0.49     & 0.09  & 2.42  & 0.07  &          \\
\\
\enddata
\end{deluxetable}

\clearpage

\begin{deluxetable}{crrrccl}
\tablenum{5}
\tabletypesize{\scriptsize}
\tablewidth{0pt}
\tablecaption{ISOPHOT PHT 22 {\it single-pointing} photometry. \label{tbl-5}}
\tablecolumns{6}
\tablehead{
\colhead{Name}                        & \colhead{${\rm \lambda}$}         &
\colhead{ Flux}                       & \colhead{$\sigma_{Flux} $}        &
\colhead{ F$_{back}$}                 & \colhead{$\sigma_{background}$}&
\colhead{Notes}     \\
\colhead{}                  & \colhead{($\mu$m)}          &
\colhead{(Jy)}             & \colhead{(Jy)}               &
\colhead{($MJy sr^{-1}$)}  &\colhead{($MJy sr^{-1}$)}     &
\colhead{}
}

\startdata

NGC 3136 ....   &  60  &$\leq $0.24&\nodata& 8.96  & 0.28  & non detection \\
                &  90  &$\leq $0.27&\nodata&13.28  & 0.54  & non detection \\
\\
NGC 3193 ....   &  60  &$\leq $0.29&\nodata&19.45  & 0.67  & non detection \\
                &  90  &$\leq $0.26&\nodata&12.31  & 0.43  & non detection \\
                & 170  &$\leq $0.34&\nodata& 6.00  & 0.19  & non detection \\
                & 200  &$\leq $0.35&\nodata& 4.06  & 0.22  & non detection \\
\\
\enddata
\end{deluxetable}

\clearpage

\begin{deluxetable}{crrrccl}
\tablenum{6}
\tabletypesize{\scriptsize}
\tablewidth{0pt}
\tablecaption{ISOPHOT PHT 22 {\it Mini-map}  photometry. \label{tbl-6}}
\tablecolumns{6}
\tablehead{
\colhead{Name}                        & \colhead{${\rm \lambda}$}         &
\colhead{ Flux}                       & \colhead{$\sigma_{Flux} $}        &
\colhead{ F$_{back}$}                 & \colhead{$\sigma_{background}$}&
\colhead{Notes}     \\
\colhead{}                  & \colhead{($\mu$m)}          &
\colhead{(Jy)}             & \colhead{(Jy)}               &
\colhead{($MJy sr^{-1}$)}  &\colhead{($MJy sr^{-1}$)}     &
\colhead{}
}

\startdata

NGC 147  ....   &  60  &$\leq $0.06&\nodata& 9.97  & 0.30  & non detection \\
                &  90  &$\leq $0.09&\nodata&10.01  & 0.18  & non detection \\
                & 170  &$\leq $0.21&\nodata&15.34  & 0.61  & non detection \\
                & 200  &$\leq $0.21&\nodata&12.63  & 0.25  & non detection \\
\\
NGC 185  ....   &  60  &  0.28     & 0.04  &10.97  & 0.33  &               \\
                &  90  &  0.92     & 0.03  &13.24  & 0.42  &               \\
                & 170  &  1.61     & 0.21  &16.37  & 0.49  &               \\
\enddata
\end{deluxetable}

\clearpage

\begin{deluxetable}{crrrrccl}
\tablenum{7}
\tabletypesize{\scriptsize}
\tablewidth{0pt}
\tablecaption{ISOPHOT P32 photometry. \label{tbl-1}}
\tablecolumns{8}
\tablehead{
\colhead{Name}                              & \colhead{${\rm \lambda}$}        &
\colhead{ F$_{model}$\tablenotemark{a}}     &
\colhead{ F$_{map}$\tablenotemark{b}}       & \colhead{$\sigma_{map}$}     &
\colhead{ F$_{back}$\tablenotemark{c}}      & \colhead{$\sigma_{background}$}   &
\colhead{Notes\tablenotemark{d}}     \\
\colhead{}                  & \colhead{($\mu$m)}          &
\colhead{(Jy)}              &
\colhead{(Jy)}              & \colhead{(Jy)}              &
\colhead{($MJy sr^{-1}$)}   &\colhead{($MJy sr^{-1}$)}                   &
\colhead{}
}
\startdata

NGC 0807 ....   & 150  & 2.53        &  2.67     & 0.09  & 13.18 & 0.43  & beam-fit  \\
\\
NGC 1172 ....   & 150  &$\leq $0.19  &$\leq $0.24&\nodata&  9.62 & 0.41  & non detection \\
                & 200  &$\leq $0.21  &$\leq $0.27&\nodata&  9.44 & 0.32  & non detection \\
\\
NGC 1344 ....   &  60  &$\leq $0.10  &$\leq $0.12&\nodata&  9.02 & 0.54  & non detection \\
                &  90  &$\leq $0.09  &$\leq $0.09&\nodata&  6.71 & 0.46  & non detection \\
                & 180  &$\leq $0.12  &$\leq $0.09&\nodata&  3.25 & 0.27  & non detection \\
\enddata
\tablenotetext{a}{The flux derived from a model fit to the data; For the nondetections the upper limits are
indicated as the flux densities to the 3 $\sigma$ uncertainties in the amplitude of the psf-fit model.}
\tablenotetext{b}{The flux density derived from direct integration of the reduced maps}
\tablenotetext{c}{The average background brightness}
\tablenotetext{d}{Beam-fit indicates that a beam model has been used
to extract the photometry for pointlike sources.}
\tablenotetext{e}{The F$_{model}$ flux is a lower limit to the true flux, since our fit
model it is not an  appropriate fit for the source. }
\end{deluxetable}

\clearpage

\begin{deluxetable}{lccrr}
\tablenum{8}
\tablewidth{0pt}
\tablecaption{ISOCAM Mid-infrared photometry}
\tablecolumns{5}
\tablehead{
\colhead{Name}                              &
\colhead{${\rm \lambda_{ref}}$}                   &
\colhead{${\rm \delta \lambda}$}            &
\colhead{Flux}                              &
\colhead{$\sigma$}                            \\
\colhead{}                  &
\colhead{($\mu$m)}          &
\colhead{($\mu$m)}          &
\colhead{(mJy)}             &
\colhead{(mJy)}
}
\startdata

NGC 185.......                 & 14.3 &    5  & 53.8 &   8.1    \\
NGC 1275 ...                   & 14.3 &    5  & 221.4 &  26.5   \\
NGC 1332 ...                   & 12.0 &    6  & 126.6 &  21.1   \\
NGC 1395 ...                   & 14.9 &    2  & 37.0  &  16.2   \\
NGC 1553 ...                   & 12.0 &    6  & 162.3 &  29.2   \\
NGC 2328 ...                   & 14.3 &    5  &  84.3 &  12.6   \\
NGC 3928 ...                   & 14.3 &    5  & 165.7 &  32.8   \\
NGC 3998 ...                   & 14.3 &    5  & 103.7 &  19.6   \\
\enddata
\tablenotetext{a}{Photometry is from Ferrari et al. (2002).}

\end{deluxetable}

\clearpage

\begin{landscape}

\begin{deluxetable}{lrrrrrrrrrrrrr}
\tablenum{9}
\tabletypesize{\scriptsize}
\tablewidth{0pt}
\tablecaption{Dust Temperatures and Masses  }
\tablecolumns{10}
\tablehead{

\colhead{}                              &
\multicolumn{4}{c}{Single-Blackbody Model, $\beta$ variable}        & 
\colhead{}                              &
\multicolumn{3}{c}{Single-Blackbody Model, $\beta$=2}           &
\colhead{}                              &
\multicolumn{4}{c}{Two-Blackbody Model, $\beta$=2}      \\
\cline{2-5}  
\cline{7-9}
\cline{11-14}\\

\colhead{Name}                          &
\colhead{reduced $\chi^2$}                      &
\colhead{$\beta$}                       &
\colhead{$T_D$}                         &
\colhead{$M_D$}                         &
\colhead{}                              &
\colhead{reduced $\chi^2$}                      &
\colhead{$T_D$}                         &
\colhead{$M_D$}                         &
\colhead{}                              &
\colhead{$T_D^{warm}$}                  &
\colhead{$M_D^{warm}$}                  &
\colhead{$T_D^{cold}$}                  &
\colhead{$M_D^{cold}$}                  \\
\colhead{}                              &
\colhead{}                              &
\colhead{}                              &
\colhead{(K)}                           &
\colhead{($\times 10^4 M_{\odot}$)}     &
\colhead{}                              &
\colhead{}                              &
\colhead{(K)}                           &
\colhead{($\times 10^4 M_{\odot}$)}     &
\colhead{}                              &
\colhead{(K)}                           &
\colhead{($\times 10^4 M_{\odot}$)}     &
\colhead{(K)}                           &
\colhead{($\times 10^4 M_{\odot}$)}     
}

\startdata

NGC 0147 .......   &\nodata&\nodata  &\nodata  &\nodata   & &\nodata&\nodata &\nodata    & &\nodata&\nodata   & 20 &$\leq$0.041 \\
NGC 0185 .......   & 1.8   & 1.2     & 26    & 0.039   & & 1.9   & 22   & 0.15     & & 42  &0.00023  & 21 & 0.16       \\
NGC 0221 .......   &\nodata&\nodata  &\nodata  &\nodata   & &\nodata&\nodata &\nodata    & &\nodata&\nodata   & 20 &$\leq $0.065 \\
NGC 0596 .......   &\nodata&\nodata  &\nodata  &\nodata   & &\nodata&\nodata &\nodata    & &\nodata&\nodata   & 20 &$\leq $62 \\
NGC 0720 .......   &\nodata&\nodata  &\nodata  &\nodata   & &\nodata&\nodata &\nodata    & &\nodata&\nodata   & 20 &$\leq $24  \\
NGC 0807 .......   &\nodata& 1.0     & 25    & 690      & & 7.4   & 20   & 3840      & & 43  &  12    & 19 & 5270       \\
NGC 0821 .......   &\nodata&\nodata  &\nodata  &\nodata   & &\nodata&\nodata &\nodata    & &\nodata&\nodata   & 20 &$\leq $61  \\
NGC 1172 .......   &\nodata&\nodata  &\nodata  &\nodata   & &\nodata&\nodata &\nodata    & &\nodata&\nodata   & 20 &$\leq $99  \\
\enddata

\tablecomments{The estimated errors on the derived temperatures are $\sim 10\%$.
}

\end{deluxetable}
\end{landscape}

\clearpage

\begin{deluxetable}{crccccccr}
\tablenum{10}
\tabletypesize{\scriptsize}
\tablewidth{0pt}
\tablecaption{FIR angular extent}
\tablecolumns{7}
\tablehead{
\colhead{Name}                              &
\colhead{ $R_e $ \tablenotemark{a}}       &
\colhead{${\rm \lambda}$}                   &
\colhead{S\tablenotemark{b}}                &
\colhead{$\sigma_{T,major}$\tablenotemark{c}} &
\colhead{$\sigma_{S,major}$\tablenotemark{d}} &
\colhead{$\sigma_{T,minor}$\tablenotemark{c}} &
\colhead{$\sigma_{S,minor}$\tablenotemark{d}} &
\colhead{ P.A.\tablenotemark{e}}            \\
\colhead{}                  &
\colhead{arcsec}            &
\colhead{($\mu$m)}          &
\colhead{($MJy sr^{-1}$)}   &
\colhead{(arcsec)}              &
\colhead{(arcsec)}              &
\colhead{(arcsec)}              &
\colhead{(arcsec)}              &
\colhead{degree}
}
\startdata

NGC 4261 .... & 38.6 & 100  & 3.62   & 37.5   & 32.5  & 22.1   & 11.9  & 1     \\
              &      & 150  & 1.12   & 60.7   & 46.0  & 42.2   & 13.7  & 59    \\
\\
NGC 5173 .... & 14.1 &  60  & 5.29   & 25.1   & 16.8  & 20.1   & 7.6   & -51   \\
              &      & 100  & 6.58   & 25.3   & 14.9  & 20.9   & 4.5   &  9    \\
              &      & 150  & 2.93   & 48.7   & 27.8  & 40.3   & 5.7   & -75   \\
              &      & 200  & 1.98   & 49.0   & 26.7  & 44.8   & 17.8  & -76   \\
\\
\enddata
\tablenotetext{a}{Optical effective radii are from \citet{fab89}. For those galaxies
not included in their list, $R_e $ has been derived from the linear relation
between $log R_e$ and $M_v$ presented in \citet{fab97}, assuming $B-V = 1$.}
\tablenotetext{b}{The amplitude S of the source model. }
\tablenotetext{c}{The width $\sigma_{T,major}$ ($\sigma_{T,minor}$) in arcsec of 
the major (minor) axis of the fitted elliptical gaussian. The FWHM is 2.354$\sigma_{T,major}$ }
\tablenotetext{d}{The width $\sigma_{S,major}$ ($\sigma_{S,minor}$) in arcsec of 
the major (minor) axis of the gaussian source model.}
\tablenotetext{e}{The Position Angle P.A. of the two-dimensional fitted gaussian.
The position angle is defined as the rotation of the ellipse from the scan direction
axis (y-axis) clockwise. }
\tablecomments{NGC 5666 at 200 $\mu$m has a complex surface
brightness; our simple two-dimensional gaussian fit
has to be considered as a temptative fit. A multi-component fit
would be more appropriate to descibe the light profile at this wavelength.}
\end{deluxetable}

\end{document}